%% file: Main.tex
\documentclass[sigconf,nonacm]{acmart}

\AtBeginDocument{%
  \providecommand\BibTeX{{%
    \normalfont B\kern-0.5em{\scshape i\kern-0.25em b}\kern-0.8em\TeX}}}

\usepackage{graphicx}
\usepackage{lipsum}
\usepackage{subcaption}
\usepackage{booktabs}
\usepackage{textcomp}
\usepackage{siunitx}
\usepackage{xcolor}
\usepackage[export]{adjustbox}
\usepackage{amsmath}
\usepackage[linesnumbered,ruled]{algorithm2e}
\usepackage[skip=2pt]{caption}
\usepackage[english, status=final, margin=false, inline=true]{fixme}
\usepackage{enumitem}
\usepackage{hyperref}
\PassOptionsToPackage{bookmarks=false}{hyperref}

\acmYear{2020}
\setcopyright{none} 
\acmConference[IPSN'20]{The 19th ACM/IEEE Conference on Information Processing in Sensor Networks}{APRIL 21-24, 2020}{Sydney, Australia}
\begin{document}

\title{Group-In: Group Inference from Wireless Traces of Mobile Devices}

\author{G{\"u}rkan Solmaz}
\affiliation{%
  \institution{NEC Laboratories Europe}}
\email{gurkan.solmaz@neclab.eu}

\author{Jonathan F{\"u}rst}
\affiliation{
  \institution{NEC Laboratories Europe}}
\email{jonathan.fuerst@neclab.eu}

\author{Samet Ayta\c{c}}
\affiliation{%
  \institution{Bo\u{g}azi\c{c}i University}}
\email{samet.aytac@boun.edu.tr}

\author{Fang-Jing Wu}
\affiliation{
 \institution{TU Dortmund University}}
\email{fang-jing.wu@tu-dortmund.de}

\renewcommand{\shortauthors}{Solmaz et al.}

\begin{abstract}

This paper proposes {\em Group-In}, a wireless scanning system to detect static or mobile people groups in indoor or outdoor environments. Group-In collects {\em only} wireless traces from the Bluetooth-enabled mobile devices for group inference. The key problem addressed in this work is to detect not only static groups but also moving groups with a multi-phased approach based only noisy wireless Received Signal Strength Indicator (RSSIs) observed by multiple wireless scanners without localization support. We propose new centralized and decentralized schemes to process the sparse and noisy wireless data, and leverage graph-based clustering techniques for group detection from short-term and long-term aspects. Group-In provides two outcomes: 1) group detection in short time intervals such as two minutes and 2) long-term linkages such as a month. To verify the performance, we conduct two experimental studies. One consists of 27 controlled scenarios in the lab environments. The other is a real-world scenario where we place Bluetooth scanners in an office environment, and employees carry beacons for more than one month. Both the controlled and real-world experiments result in high accuracy group detection in short time intervals and sampling liberties in terms of the Jaccard index and pairwise similarity coefficient.

\end{abstract}

\maketitle
\footnotesize \copyright 2020 IEEE. Personal use of this material is permitted. Permission from IEEE must be obtained for all other uses, in any current or future media, including reprinting/republishing this material for advertising or promotional purposes, creating new collective works, for resale or redistribution to servers or lists, or reuse of any copyrighted component of this work in other works.

\normalsize

\input{Introduction.tex}

\input{RelatedWork.tex}
\input{Approach.tex}

\input{Experiment.tex}

\input{Conclusion.tex}


\bibliographystyle{ACM-Reference-Format}
\balance
\bibliography{Main}
\end{document}

%% file: Introduction.tex
\begin{figure}
\centering
  \begin{subfigure}[b]{.23\textwidth}
\includegraphics[width=0.95\textwidth]{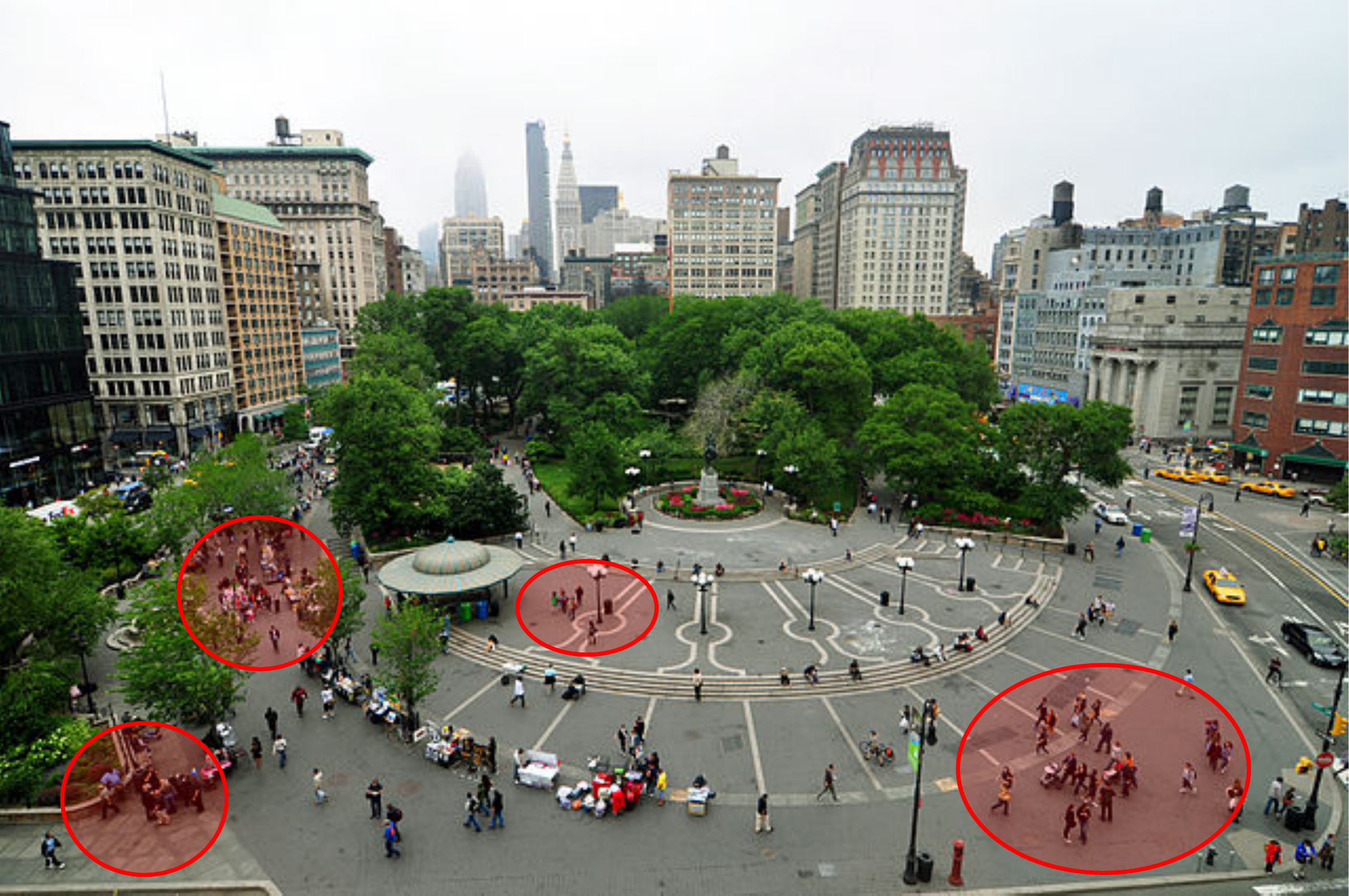}
\caption{Finding groups in urban areas.}
\end{subfigure}~
\begin{subfigure}[b]{0.235\textwidth}
\centering
  \includegraphics[width=1\textwidth]{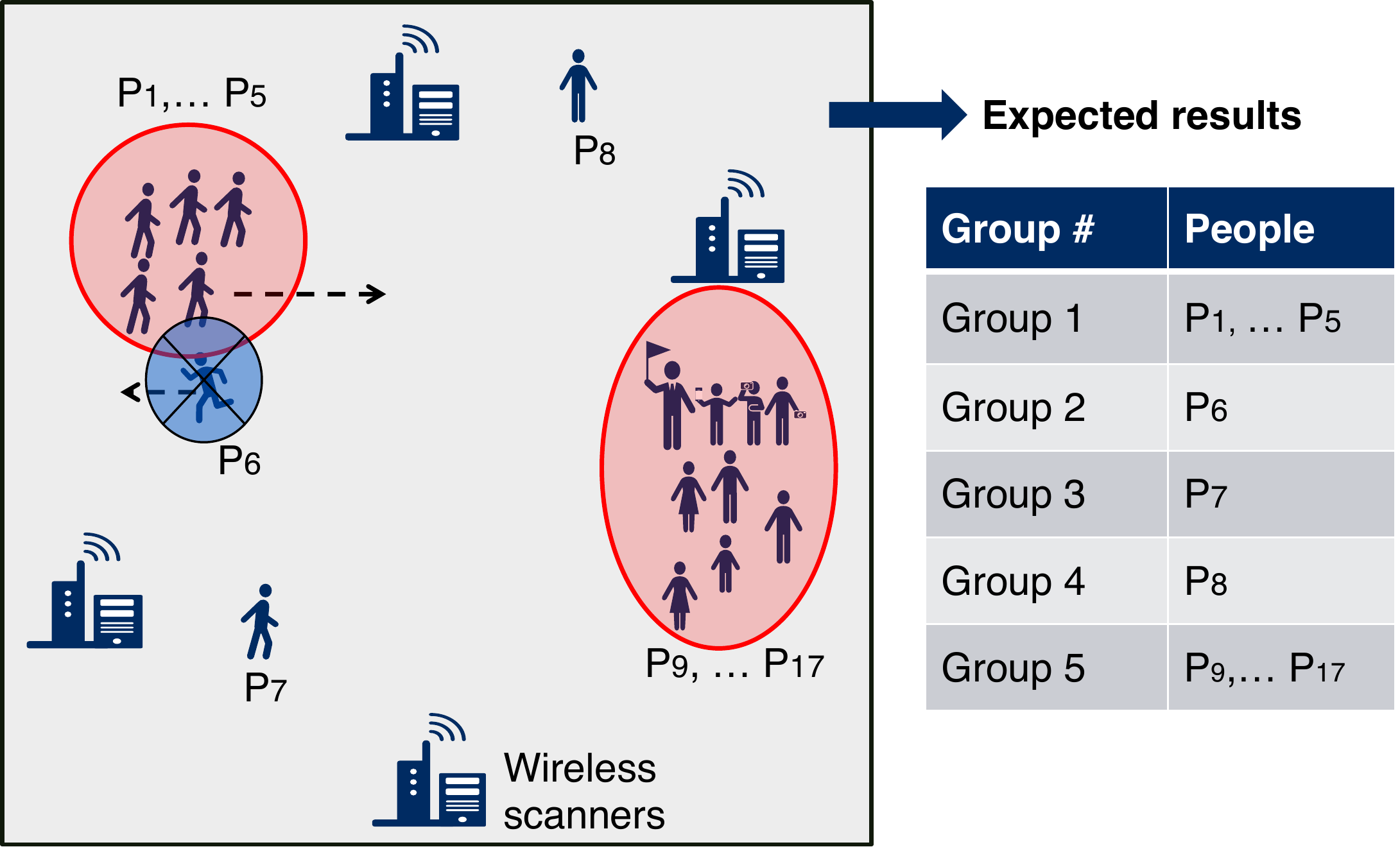}
\caption{Group detection expected outputs.} 
\label{Fig:GroupingGoal}
\end{subfigure}
\caption{Inferring static and mobile groups using wireless scanners.}
\end{figure}
\section{Introduction}

There has been increasing interest in recognizing~\cite{guo2017wifi,wang2015understanding} human mobility behaviors to create smarter future environments. In particular, group detection is beneficial for various domains. For instance, accurate real-time and offline group detection can be helpful in the following scenarios. 

\begin{itemize}[leftmargin=*]
\item \textbf{Crowd management:} The group mobility behaviors heavily affect crowd dynamics~\cite{moussaid2010walking}. Smart cities develop strategies based on knowledge of groups for improved congestion avoidance, evacuation planning, and demand management. 

\item \textbf{Retail scenarios}: Retailers can promote their products based on groups in shopping malls, as suggested in~\cite{sen2014grumon}. Understanding customer profiles (e.g., singles or couples) results in improved recommendation systems.

\item \textbf{Evacuation modeling:} As movements of people in groups~\cite{moussaid2010walking} are affected by group interests rather than individuals' movement decisions; simulations can leverage data provided by group detection.

\item \textbf{Social isolation detection:} Group detection can be useful to analyze social engagement or isolation. For instance,  to monitor the elderly in assisted living places for their interactions~\cite{guo2017wifi}.
\end{itemize}

Different than the typical proximity-based group detection, we regard ``group" in this work as an indicator of people spending time together. For example, people are sitting together for a tea break (and thus forming a static group), or people are visiting a place by walking together (forming a moving group).

Most of the existing video-based or wireless approaches consider high-accuracy localization~\cite{ge2012vision}, which may require a vast amount of data collection during calibration, training, and operation phases. Other methods that leverage wireless access points mostly suffer from long coverage ranges. For instance, being in the same hotspot area is considered as an encounter~\cite{larsen2013crowds,kostakos2010brief}. On the other hand, long ranges such as 100m cause coarse granularity. Lastly, some approaches rely on mobile sensing data collection~\cite{sen2014grumon}. These approaches depend on existing user incentives for active usage of apps as well as battery and data consumption. Group-In does not require high-accuracy localization, smartphone information, whereas it still provides fine-grained group inference.

We propose the {\em Group-In} system for finding groups using only {\em wireless traces} from people's mobile devices, as shown in Fig.~\ref{Fig:GroupingGoal}. The system leverages a distributed wireless data collection mechanism with multiple scanners. The scanners detect different mobile devices in their vicinity and extract the RSSI values, which are estimated measures for the power level of the received signals by the scanners. The system uses these measures and accurately detects the existence of groups in shorter (e.g., in a city square) or more extended periods (e.g., a working week at an office where people spend most of their day). Group-In is applicable in a wide range of scenarios. The proposed design can be used in small environments such as a room, whereas one can scale the system to larger environments (e.g., city-scale). Furthermore, Group-In does not require a data collection or environment-specific calibration phase, while neither relying on expensive and potentially privacy-invading systems that depend on camera feeds. Small changes in configurations (such as the scanner locations) do not affect Group-In algorithms. We observe that Group-In produces accurate results in many different lab setups (scanner placement schemes, group mobility behaviors) using the same parameters for the algorithms without any configuration change. Thus, it is a very flexible and easy-to-use system that is useful in temporary (e.g., festivals) or permanent (e.g., office rooms) setups without much effort of following strict deployment, configuration, or testing schemes.

There exist two key technical challenges for accurately detecting both static and dynamic moving groups: (1) {\em processing sparse and noisy wireless data} and (2) {\em combining data from multiple scanners} in the environment. The first challenge arises from the nature of radio-frequency (RF) signals, mainly for reasons such as wireless data loss due to interference, scanners missing the Bluetooth (BT) advertisement packets, and the inactivity of user devices to preserve battery power. The second challenge arises due to scanners having partial views of their environments with limited wireless coverage. Data sparsity results in different and changing numbers of dimensions for the wireless trajectories of mobile devices. Thus, one cannot merely compare the trajectory distances or apply clustering algorithms to achieve high accuracy group detection.

This paper proposes a step-by-step approach that consists of the wireless scanning with multiple observers, preprocessing steps, {\em centralized and decentralized analytics}, {\em group detection}, and {\em long-term linkage} analysis. We propose new analytics algorithms considering static/mobile group detection aspects. The centralized computing approach aggregates the RSSI data in the back-end server and creates a graph using pairwise distances of wireless trajectories. In decentralized computing, each scanner creates its results through wireless trajectory comparisons, and later these results are unified. Both centralized and decentralized computing can analyze data from different mobile devices with varying numbers of scanners data and accurately classify groups, even if the data is sparse and noisy. Centralized computing provides higher accuracy group monitoring, whereas decentralized computing enables keeping the collected data in the devices and collecting only the result messages of the scanners in the server. Centralized setups of Group-In can be beneficial when the system operates in wireless local area networks controlled by a wireless provider such as a university or elderly care center where registered people can be traced accurately. Decentralized setups enable large-scale scenarios where many scanners collect the data, or bandwidth usage is expensive/limited. For applications such as city-wide monitoring, the decentralized setup enables most of the processing of many wireless packets on the devices and Group-In can still provide accurate group monitoring in large-scale. In the {\em group detection} step, Group-In feeds the outputs of the algorithms to the graph clustering algorithms. In the last step, Group-In finds {\em long-term linkages} by aggregating the group detection outputs over a more extended period. The long-term linkage outputs can be given as inputs for social studies.

%% file: RelatedWork.tex
\section{Related Work}
\label{sec:related-work}
Most studies related to human mobility analytics (more specifically group detection) belong to one of the three categories: 1) video-based detection, 2) wireless activity-based detection, 3) detection using data from smartphone apps or social networks. We regard data from smartphone apps and social networks as {\em user data} as these applications require user-specific data collection. In contrast, the first two categories do not require people to download an app or sign up for a service. On the other hand, user data availability allows applications beyond group detection, such as social interaction analytics or friendship detection. In this section, let us discuss some of the recent significant advancements.

\noindent
\textbf{Video-based group detection:} There have been various studies leveraging video footages to extract crowd information, and some focus on detecting groups. Ge et al.~\cite{ge2012vision} and Solera et al.~\cite{solera2016socially} propose detecting groups by clustering movement trajectories extracted from video footage (hierarchical and correlation clustering). Moussa{\"\i}d et al.~\cite{moussaid2010walking} analyze the effects of group behaviors of pedestrians to crowd dynamics. Their analysis for pedestrians observed by cameras in a commercial street shows that up to 70\% of people are moving in groups, including couples, families, or friends, and the group sizes follow a Poisson distribution. Since video-based crowd behavior learning needs labeled video datasets, the study in~\cite{cheung2016lcrowdv} aims to generate synthetic labels and combine them with real videos. While Group-In also has a clustering-based approach, it does not require camera deployment or training using labeled video datasets.

\noindent
\textbf{Wireless activity-based group detection:} There exist recent studies related to ``device-less'' wireless detection of people in indoor environments. Guo et al.~\cite{guo2017wifi} propose an approach to find the existence of people and estimating the density of people in a room, walking speed, and direction based on Wi-Fi channel state information (CSI). CSI-based approaches analyze channel properties of the communication links which are affected by the vicinity of people or objects. Their approach leverages semi-supervised learning for environments such as rooms in assisted living places. Adib et al.~\cite{adib2015multi} propose the WiTrack2.0 system for tracking moving or static users with up to 10m range for indoor localization using wireless signal reflections. The Freesense system by Xin et al.~\cite{Xin:2018:FRA:3279953.3264953} performs indoor detection by identifying phase differences between the amplitude waveforms of multiple antennas. The CrowdProbe system by Hong et al.~\cite{Hong:2018:CNC:3279953.3264925} obtains Wi-Fi probe requests and performs a Hidden Markov Model-based algorithm to detect crowd movement in a multi-floor museum.

Larsen et al.~\cite{larsen2013crowds} analyze data from BT scanners during a music festival to detect groups and understand the overall network structure. Kostakos et al.~\cite{kostakos2010brief} analyze the encounters between people in urban areas. They deploy BT scanners in Bath, UK, and analyze brief vs. persistent encounters in the city. Both of the approaches by Larsen et al. and Kostakos et al. consider sparse deployment of scanners and assume that people might be in the group when the same scanner scans them. Weppner et al.~\cite{weppner2016monitoring} collect Wi-Fi and BT data with a subset of video data to analyze group density and flow by calibrating estimations with a few ground truth points. The majority of localization estimates (90\%) are between  5 and 11~m of their ground truth location. Jamil et al.~\cite{jamil2015hybrid} use BT data to analyze groups during the Hajj pilgrimage with a hybrid participatory approach combining GPS data from few, ``leader'' smartphones and Bluetooth Low Energy (BLE) beacon data from the majority of participants. Brockmann et al.~\cite{Brockmann:2018:RBP:3234847.3234861} perform the detection of persons in a queue using BLE data. Solmaz and Wu~\cite{Solmaz-2017-ICC} propose detecting group and individual walking behaviors using BT scanners. The detection is limited to the movement detection (i.e., excluding waiting times or static groups) where people need to follow a particular path and move directly from one scanner to another.

Our approach contributes to wireless-activity based group detection. Most of the studies above are limited to either specified areas where groups need to have a specific movement behavior or groups are limited to a restricted area. On the other hand, Group-In can be used in various indoor/outdoor areas for stream-based (near-real-time) detection without specific environmental limitations. Moreover, Group-In can differentiate people who belong to different groups while being scanned by the same scanner(s). 

\noindent
\textbf{User data-driven group detection:} The GruMon system by Sen et al.~~\cite{sen2014grumon} uses smartphone sensors for tracking group behaviors based on the fusion of accelerometer, barometer and Wi-Fi location (extracted based on Wi-Fi access point locations). The detection time can be as short as 5 to 10~min. Jayarajah et al.~\cite{jayarajah2015need} apply GruMon to understand the effects of the groups on people's behaviors, including mobility patterns, responsiveness to phone calls, and app usage. Kaur et al.~\cite{Kaur:2018:SIR:3276774.3276786} collect Wi-Fi data and web query logs of users in a large shopping center to map semantic similarity across cyber and physical behaviors for future location prediction. Sonta et al.~\cite{Sonta:2018:IOT:3276774.3276779} use plug load energy sensor data to detect the social network of occupants in a commercial building. Yu and Han~\cite{7035660} propose the Grace mobile app for iOS platforms, which recognizes proximity between two devices using BT RSSI. Their approach recognizes groups when the group members are well separated from others and face to face with each other.

Sociophone~\cite{lee2013sociophone} and SocialWeaver~\cite{luo2013socialweaver} are smartphone applications to track interactions (conversations) for deeper social analysis. Canzian and Musolesi propose Mood-Traces~\cite{canzian2015trajectories}, which aims to detect depression states based on social group interactions using smartphones' GPS data. Mehrotra and Musolesi~\cite{mehrotra2018using} extend this study by automatically extracting mobility features using a deep autoencoder. D'Silva et al.~\cite{d2018role} analyze the role of human mobility dynamics in the retail business survival in cities using transportation data and crowd-sourced data from location-based social networks. Yuan et al.~\cite{yuan2017pred} extract human mobility patterns by leveraging social media data. The proposed approach can be useful for location recommendation services, such as services for local event recommendations. Yu et al.~\cite{Yu:2018:IMR:3279953.3264957} analyze user location data to construct a user graph based on their spatial-temporal interactions and learn user representations from the graph. Du et al.~\cite{Du16} combine mobile sensing using smartphones with Wi-Fi signals to detect moving, static groups, and their structures such as pairwise leader/follower relationship. Jiang et al.~\cite{jiang2017activity} propose activity-based human mobility detection focusing on tour patterns and trip-chaining behaviors with call detail record (CDR) data. There have been various studies related to the usage of CDRs. Although the localization resolution of these datasets is rather low, they are considered for city planning and optimizing transportation services.

Indoor localization techniques, including wireless triangulation or CSI, require extensive data collection and environment-specific calibration~\cite{vo2016survey}. The proposed Group-In approach explores the possibilities without trilateration or location extraction, assuming accurate people localization is not available. On the other hand, the proposed schemes in this paper could apply to the data coming from different sensors (e.g., GPS). 

Group-In uses only the information coming from wireless scanners, and it does not require users to provide data or sign up for a social network service. Furthermore, Group-In can also function as a stream-based service and produce outputs at short time intervals. The basic requirements of Group-In do not include the active participation of people or data collection from social media or telecommunication service providers. Lastly, Group-In can be scaled to large areas (e.g., smart cities) by the deployment of low-cost wireless scanners. 

Studies mentioned above require high accuracy localization of individuals through GPS and cameras, or they make coarse assumptions for groups such as being connected to the same Wi-Fi hotspot as a direct indication of groups. Group-In provides group inference with high accuracy and spatiotemporal resolution without these requirements or assumptions.

%% file: Approach.tex
\begin{figure*}[t!]
    \centering
    \begin{subfigure}[b]{0.33\textwidth}
        \includegraphics[width=\textwidth]{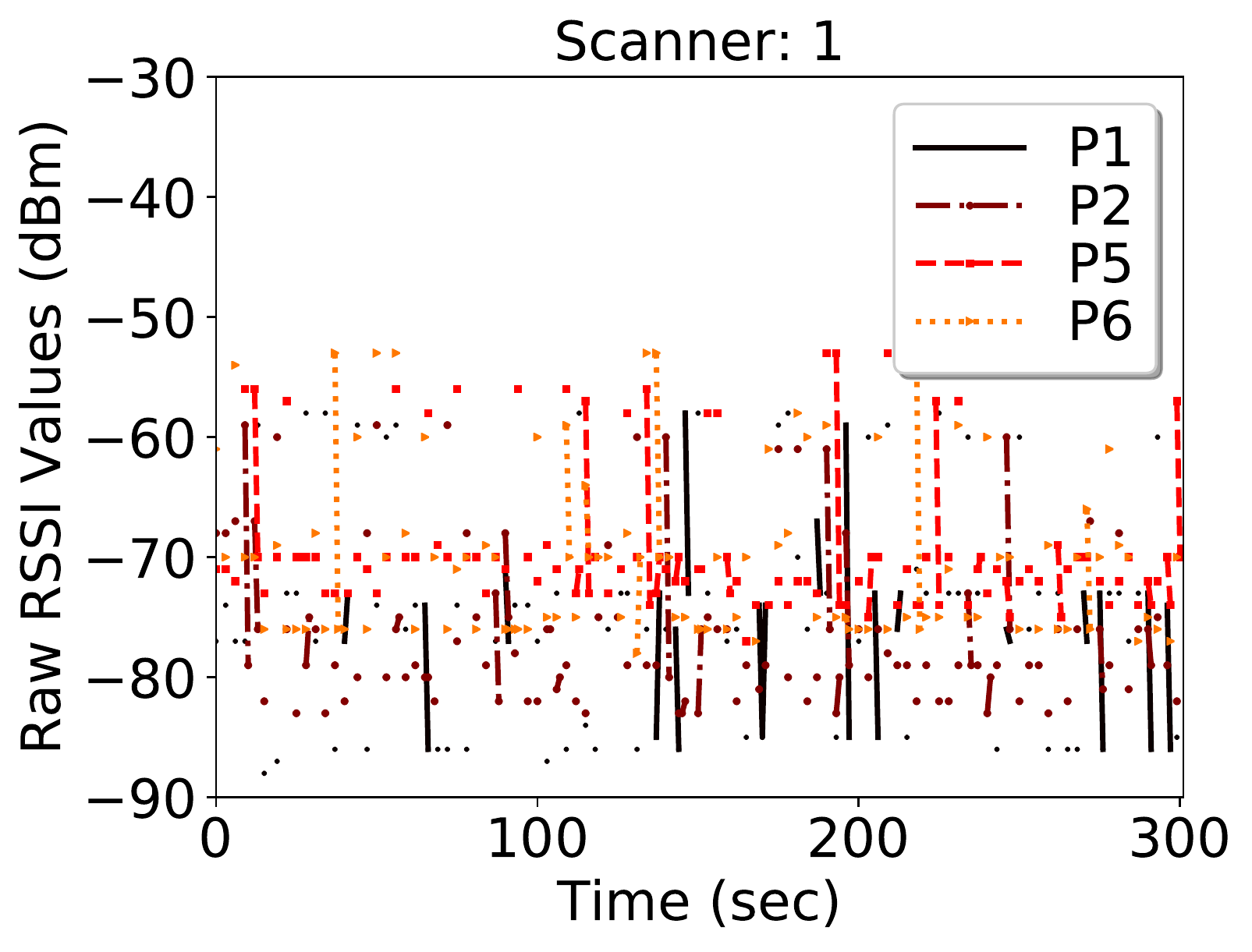}
    \end{subfigure}
    ~
    \begin{subfigure}[b]{0.33\textwidth}
        \includegraphics[width=\textwidth]{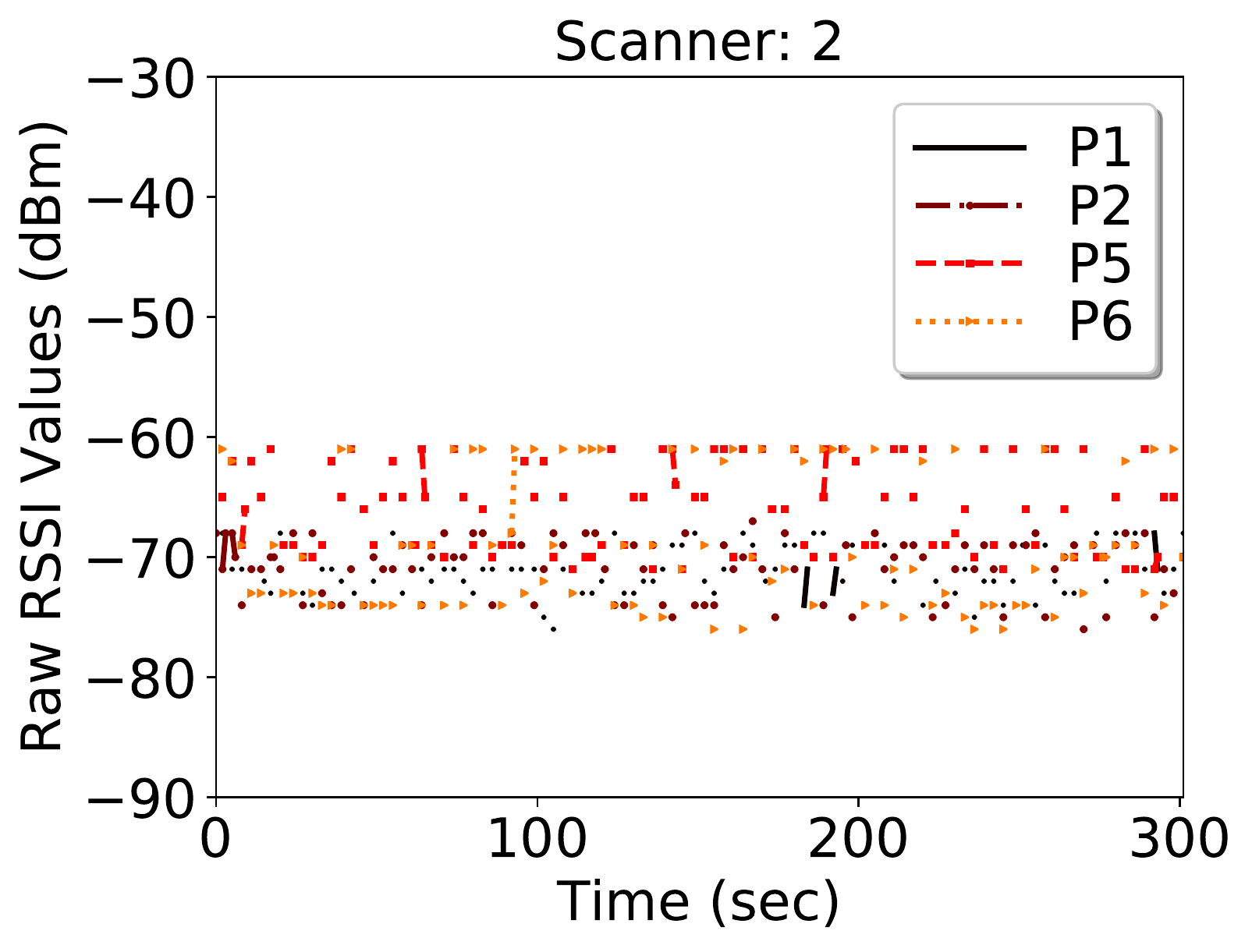}
    \end{subfigure}
    ~ 
    \begin{subfigure}[b]{0.33\textwidth}
        \includegraphics[width=\textwidth]{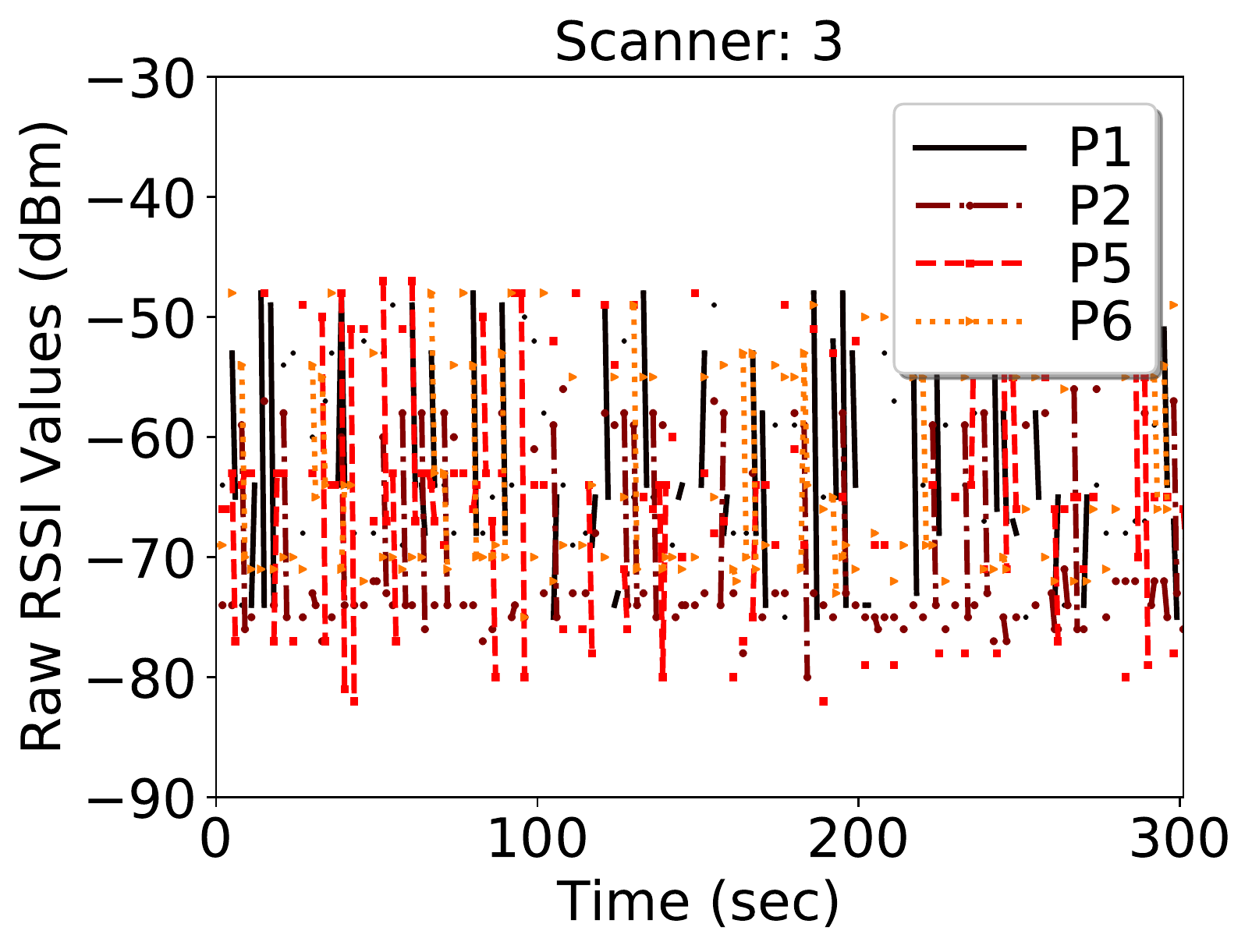}
    \end{subfigure}
    \caption{Raw RSSIs during the controlled experiments for three scanners detecting four beacons ($P_1,P_2,P_5,P_6$). 
    }\label{fig:signals}
\end{figure*}
\begin{figure*}[t!]
\centering
      \begin{subfigure}[b]{0.325\textwidth}
        \includegraphics[width=\textwidth]{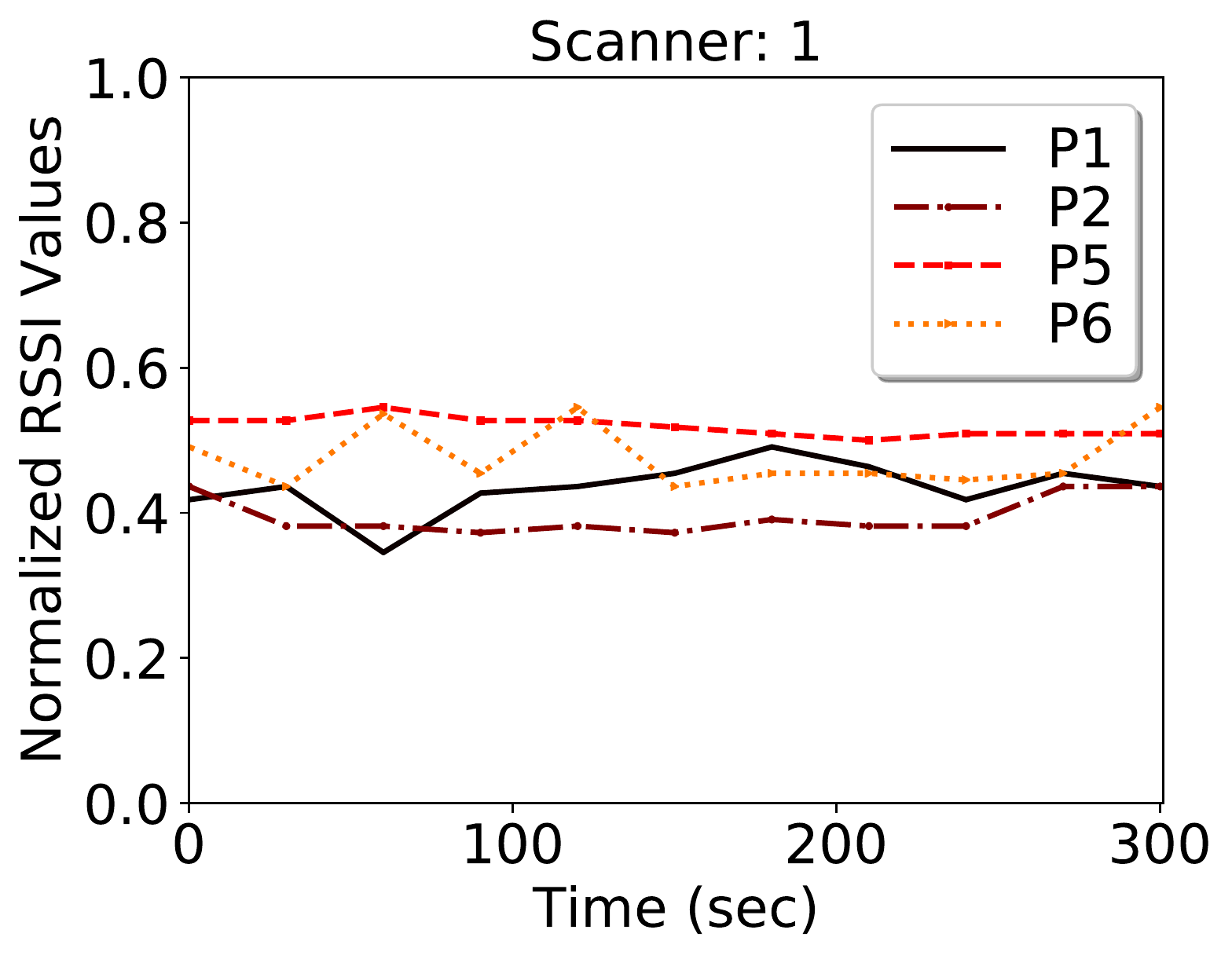}
    \end{subfigure}
    ~
    \begin{subfigure}[b]{0.325\textwidth}
        \includegraphics[width=\textwidth]{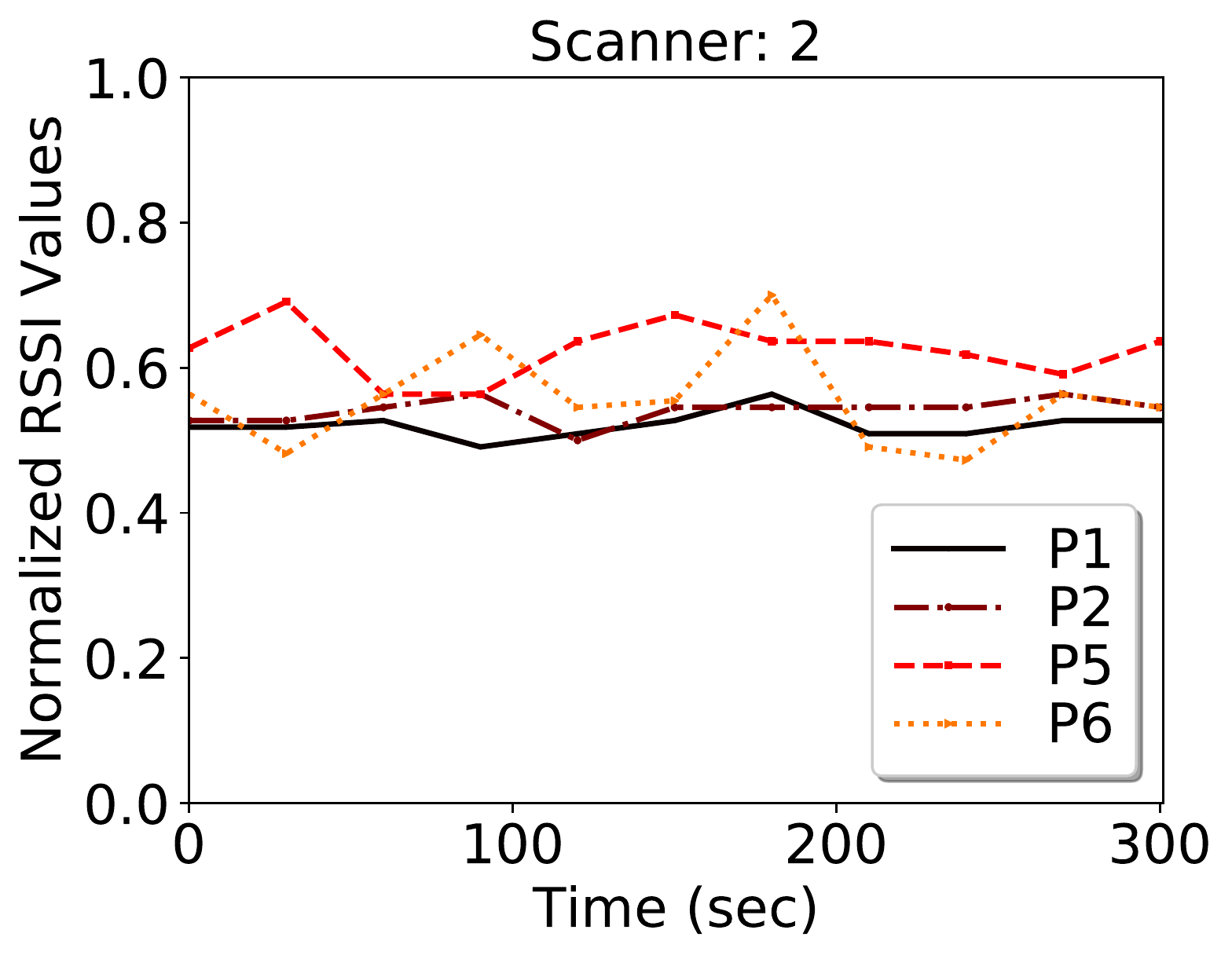}
    \end{subfigure}
    ~ 
    \begin{subfigure}[b]{0.325\textwidth}
        \includegraphics[width=\textwidth]{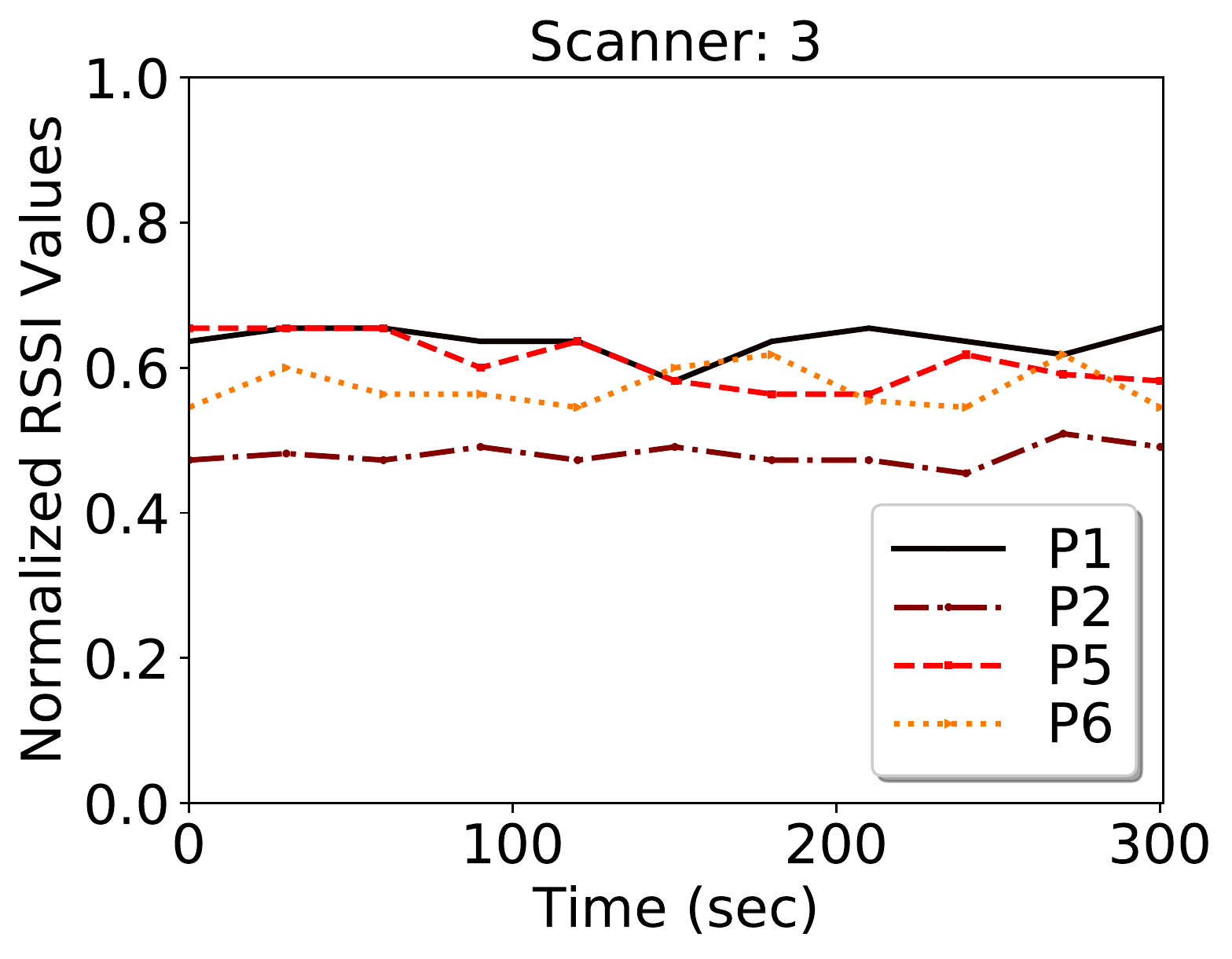}
    \end{subfigure}
        \caption{Sampled (with 30~sec) and normalized RSSIs during controlled experiments for three scanners detecting four beacons ($P_1,P_2,P_5,P_6$).
        }\label{fig:signals_sampled}
\end{figure*}

\section{Group Detection Problem}
\label{Problem}

There exist two main challenges for accurately detecting groups from wireless traces: (1) Processing sparse and noisy data and (2) combining data from multiple scanners.
\subsection{Sparse and Noisy Data}

We make our initial observations in the lab environment where three BT scanners are at the corners of a conference room (of size $100$~m$^2$) and two groups of BLE beacons (simulating people, where $P$ denotes a person) are in two different places in the room. The distance between the two static people groups is 6~m. Fig.~\ref{fig:signals} shows the raw RSSI values received by each of the scanners from 4 beacons where $P_1$ and $P_2$ belong to one group and $P_5$ and $P_6$ belong to another. This figure provides an initial perspective on the problem of group detection due to the visibly noisy and sparse nature of the raw values. 

In Fig.~\ref{fig:signals}, there is a line between two consecutive seconds, both of which have measurements from a beacon. We observe that the lines rarely appear (especially for Scanner 1 and 2), meaning the scanners do not receive packets in most of the seconds. The reason can be the channel-hopping scanners that miss the BT advertising channels.
Furthermore, even in the case of scanners have measurements from all four beacons (beacons are all in the wireless range), their observation data is not only sparse but also has different densities. For instance, the data from Scanner 3 has a higher density compared to the data from Scanner 2. Lastly, the RSSI values are noisy, as previously explicitly observed for BT signals~\cite{faragher2015location}.

\subsection{Combining Data from Multiple Scanners}

One major aspect to consider is the mobility of the groups. As a simple example, the movement of a person who goes from point X to point Y is different from the movement of another person who goes from point Y to point X during the same time interval. Fig.~\ref{Fig:GroupingGoal} shows the expected results of the group detection with a simple illustration. In this figure, people $P_1,\ldots, P_5$ are in the same group, moving in the trajectory shown with the arrow, whereas $P_6$ should not be listed in the same group. She is momentarily very close to the group, although people's movement trajectories suggest otherwise. In addition, there exists a static group with the people $P_9,\ldots,P_{17}$. Multiple wireless scanners are in the vicinity. Although not illustrated in this figure, each scanner has a wireless range (e.g., 30~m), and people may enter in or move out of the range of a scanner during their movement.

To capture the difference between movement trajectories, an accurate system needs to divide expected the time interval of group detection into shorter discrete time frames (e.g., 5~sec), and consider the difference between wireless traces for each time frame. Later, the comparison results stand for the overall time interval. Second, the intuitive approach of clustering directly on the time frame data does not solve the group detection problem since this approach assumes that the scanners detect the people at every time frame. In practice, this is not the case, as some of the devices are out of the scanners' wireless range. For an observed mobile device, one may consider the wireless data for this device from each scanner as another dimension for the data collected from the device. If we imagine all devices' data in such multi-dimensional space, inferring groups, or even comparing two people's trajectories is not a straight-forward problem since the traces may have different (number of) dimensions. The proposed approach aims to tackle these challenges.

\section{Group-In Approach}
\label{Approach}

\begin{figure*}[t!]
  \includegraphics[width=\textwidth]{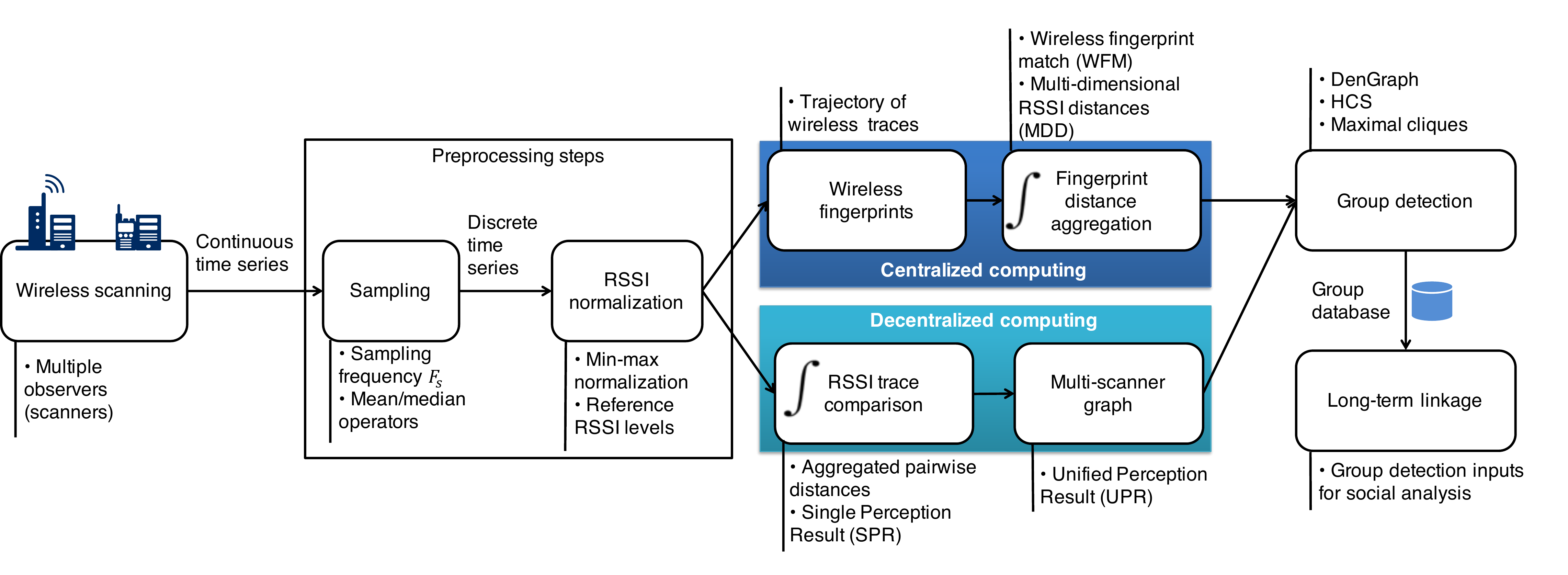}
\centering
\caption{A functional view of the Group-In approach having processing steps starting from the left (wireless scanning) and ends on the right (group detection and long-term linkage).} \label{Fig:Approach}
\end{figure*}

Fig.~\ref{Fig:Approach} shows the overall functional view of the approach to solve the group detection problem through a list of sub-tasks (steps) shown as separate boxes. While centralized computing and decentralized computing share a set of these sub-tasks, they differ in particular for pairwise comparisons. Moreover, the physical place of the computation differs, such that in centralized computing, scanners send continuous time-series data directly to the back-end server. For decentralized computing, the scanners can perform the preprocessing steps and RSSI trace comparison themselves. Let us now describe each step, starting from wireless scanning (left) to long-term linkage (right).

\subsection{Wireless Scanning with Multiple Observers}

In the initial phase of our approach, multiple wireless scanners receive wireless packets from people's mobile devices. The scanners do not necessarily cover all devices of the people in the surrounding.

\begin{figure}[t!]
\includegraphics[width=1\columnwidth]{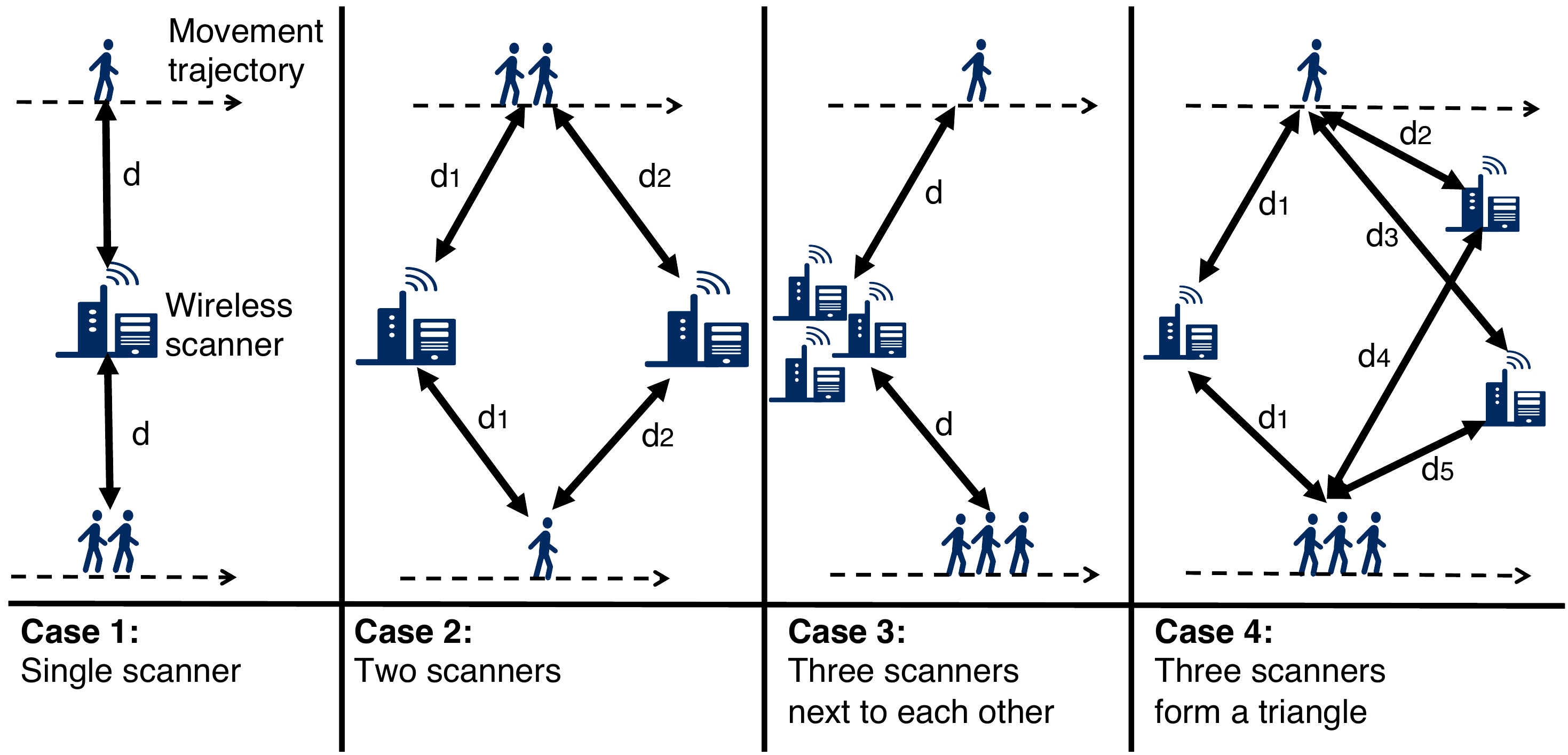}
\centering
\caption{Wireless scanner deployment scenarios.} \label{Fig:Deployment-scenarious}
\end{figure}

Fig.~\ref{Fig:Deployment-scenarious} shows four simple example wireless scanner scenarios, which may lead to success or failure for the group detection. In all cases, we consider two groups walking in parallel with the shown distances from the wireless scanners. Case 1 illustrates why Group-In needs multiple scanners such that the single scanner may not capture the difference between the person (single-person group) and the two-people group as they have both distance $d$ to the scanner. Considering the groups move through the trajectories in parallel, using a single scanner may result in a group consisting of three people over the time interval. As shown in Case 2, two scanners may also lead to failure as the two groups' distances to both scanners always remain similar. There exist three scanners in Case 3. However, their locations are next to each other. In this case, each scanner may provide similar RSSI levels, and the combination of the observations of three scanners may result in a single group.
On the other hand, as shown in Case 4, if the scanners have a certain distance to each other, they might detect the groups correctly as the distances $d_2-d_4$ and $d_3-d_5$ are distinct. Case 1 practically represents a sparse deployment as there exists only one scanner in the vicinity of the two groups (although there may be scanners far from the area, whereas Case 3 represents a very dense deployment where the distances between the scanners themselves become negligible. Deploying more scanners are necessary in this case. Hence, both sparse and very dense deployments with less than three scanners may lead to inaccuracy or inefficiency. Although Group-In does not localize mobile devices or apply triangulation (as the collected data can have any number of dimensions), its performance may decrease if the deployment of scanners is too sparse or dense. This problem also exists for typical localization approaches~\cite{7997097}. Although there is no single global scheme reached for optimal node placement, various methods exist in the literature~\cite{optimalsnifferdeployment}.

In the Group-In system, the wireless scanners constantly search for wireless packets from mobile devices in the vicinity. The received packets have a respective identifier for the device (i.e., device ID), which is captured and hashed by the scanner. The hashing provides anonymity to the device ID (e.g., MAC address of a smartphone). Along with the device ID, the scanner can extract RSSI levels and (in some cases) reference RSSI values. Reference RSSI is defined as the expected RSSI level when the device is 1~m away from the scanner. The scanner stores the extracted information from the device with the following format which we name as the {\em wireless packet} ($WP$):
\begin{align}
WP=<Time,RSSI,RefRSSI,DeviceID,ScannerID>\end{align}

\subsection{Preprocessing Steps}

Let us briefly describe the preprocessing steps, which are necessary to make the system function as expected. Current Group-In implementation uses some of the well-known methods for sampling and normalization.

\noindent
\textbf{Sampling:} We define a sampling time $\Delta t$ which is a fixed short time frame parameter (e.g., 1, 5, 10, 30, or 60~sec). For each unique pair $(DeviceID,ScannerID)$, all RSSI values of $WP$s belonging to $\Delta t$ are sampled. The RSSI values represent a continuous time series data from $RSSI_1$ to $RSSI_n$. Sampling can be performed using median or mean operators. By default we use median: 
\begin{align}
MR_{P_{j}}^{s_i}(\Delta t)=\mathcal{M}\{(RSSI_1), (RSSI_2), \ldots, (RSSI_n)\},\\
MRR_{P_{j}}^{s_i}(\Delta t)=\mathcal{M}\{(RefRSSI_1),\ldots,(RefRSSI_n)\},
\end{align}
where $MR$ and $MRR$ denote median RSSI and median reference RSSI respectively. $s_i$ denotes a scanner with index $i$ and $P_j$ denotes the device of the person with index $j$ who is identified by $DeviceID$. The output of the sampling step is the discrete time series data that is used in the following steps.

\noindent
\textbf{RSSI normalization:}
Different devices often show varying RSSI characteristics depending on the used radio chip, amplifier, antenna, and case~\cite{furst2018evaluating}. For our experiments, we leverage simple heuristics of min-max normalization (on a scale of 0 and 1) and the $RefRSSI$ contained in the Bluetooth advertisement packets for the BLE beacons. For each beacon, the simple usage of $RefRSSI$ consists of having a global average (expected) $RefRSSI$, comparing with $RefRSSI$ of the measurement, and adjusting (shifting) the corresponding $RSSI$ based on this difference. For cases when the devices are heterogeneous, and $RefRSSI$ is not available, different approaches, such as the method presented in~\cite{kjaergaard2011indoor} based on RSSI ratios or a device model database with $RefRSSI$ can be applied.

\noindent
\textbf{Preprocessing outputs:}
Outputs of the RSSI normalization step are a set of {\em wireless traces}. We define a wireless trace $WT$ as the following triple:
\begin{align}
\begin{centering}
WT(\Delta t)=(s_i, P_j, \mathcal{N}(MR^{s_i}_{P_{j}}(\Delta t),MRR^{s_i}_{P_{j}}(\Delta t))\big),
\end{centering}
\end{align}
where $\mathcal{N}$ denotes the normalization function using $MR$ (and if applicable $MRR$). A wireless trace packet is denoted as $WT^{s_i}_{P_j}$ given scanner $s_i$ and device of the person $P_j$. $WT(\Delta t).scanner=s_i$, $WT(\Delta t).person=P_j$, $WT^{s_i}_{P_{j}}(\Delta t).NRSSI$ denote the three selectors for scanner ID, person ID, and the normalized RSSI respectively.
\noindent
\textbf{Initial observation:}
Fig.~\ref{fig:signals_sampled} is the observed result after these preprocessing steps (compare to the raw RSSIs in Fig.~\ref{fig:signals}). We observe that the measurements of the different devices become more apparent compared to the raw measurements. The values from $P_1$ and $P_2$ seem similar in Scanner 1 and 2, whereas they are distinct for Scanner 3. On the other hand, the two groups ($P_1$-$P_2$ and $P_5$-$P_6$) are still not visibly different.

\subsection{Centralized Computing}

Centralized computing begins with creation of {\em wireless fingerprints} ($WF$) which are trajectories of wireless traces. For a given {\em time interval} $T$, $WF$ is given by:
\begin{equation}
\begin{split}
WF_{P_j}(T) = \bigg\{ \Big( WT_{P_j}[1](\Delta t_1), \ldots,WT_{P_j}[k](\Delta t_1)\Big), \\
 \Big( WT_{P_j}[1](\Delta t_2), WT_{P_j}[2](\Delta t_2),\ldots, WT_{P_j}[l](\Delta t_2)\Big), \\
 \ldots\\
\Big( WT_{P_j}[1](\Delta t_x), WT_{P_j}[2](\Delta t_x),\ldots,WT_{P_j}[m](\Delta t_x)\Big)
\bigg\},
\end{split}
\label{eq:WF}
\end{equation}
where $0\leq k,l,m \leq n$ and $n$ is the number of scanners. $T$ consists of a set of sampling times ($\Delta t)$, where $T=\{\Delta t_1, \Delta t_2,\ldots,\Delta t_x \}$. Here, we define an ordered list of wireless traces for every sampling time $\Delta t$ based on the normalized RSSI values as follows:
\begin{equation}
\label{eq:orderedlist2}
WT_{P_j}[1](\Delta t).NRSSI \geq WT_{P_j}[2](\Delta t).NRSSI\geq \ldots
\end{equation}
such that $WT_{P_j}[1](\Delta t)$ points to the wireless trace with highest $NRSSI$. $WF$ is a set of these ordered lists for all $\Delta t \in T$. The ordered list can be empty, meaning that there is no observation for $P_j$ at the particular sampling time. $WF_{P_j}(\Delta t) \subseteq WF_{P_j}(T)$ denotes the list of $WT$s for a given sampling time (a row in Eq.~\ref{eq:WF}).

For fingerprint distance aggregation step, the first algorithm we propose is called {\em \textbf{M}ulti-\textbf{D}imensional RSSI \textbf{D}istances} ($MDD$). Algorithm~\ref{Algo:MultiDim} computes the distance (shifted in linear proximity domain based on the log-distance path loss model~\cite{chintalapudi2010indoor}) between two people's $WF$s for a sampling time $\Delta t$ in the existence of lacking scanner data (i.e., the different number of dimensions). In this algorithm, the first iteration computes the maximum possible distance between $NRSSI$s considering the number of scanners that are present for the sampling time $\Delta t$, which corresponds to the number of dimensions. The second iteration computes the distance in this multi-dimensional space. For each dimension $i$, following possibilities exist: both $P_a$ and $P_b$ are observed by $s_i$, only $P_a$ is observed, or only $P_b$ is observed. Based on these possibilities, iteration aggregates the shifted $NRSSI$ differences. The algorithm has a parameter $\zeta$, which is empirically set as the maximum possible distance based on dimensions ($10^0 < \zeta \leq 10^1$). The output of this algorithm is the pairwise distance (denoted as $\Phi_{ab}(\Delta t)$) for the sampling time. Through the aggregation of all sampling times of a time interval $T$, the following equation calculates $MDD$ for $T$ as follows:

\begin{equation}
      MDD_{ab}(T) = \frac{\sum _{\Delta t}^{\Gamma_{ab}(T)} \Phi_{ab}(\Delta t)}{\left |\Gamma_{ab}(T)\right | } \text{ for } a \neq b, 
\end{equation}

$\Gamma_{ab}(T)$ is the set of sampling times ($\Delta t \in T$), where at least one of $P_a$ and $P_b$ has $WF(\Delta t)$ (i.e., scanners observe at least one of them even though there is no observation for the other person).

\begin{algorithm}
\footnotesize
\caption{Multi-dimensional RSSI distance} 
\label{Algo:MultiDim}
\SetKwInOut{Input}{Input}
\SetKwInOut{Output}{Output}
\Input{$WF_{P_{a}}(\Delta t)$,$WF_{P_{b}}(\Delta t)$: Wireless fingerprints}
\Input{$n$: Number of scanners during $T$}
\Input{$\zeta$ (parameter): Max possible distance}
\Output{$\Phi_{ab}(\Delta t)$: Pairwise distance.}
$\psi \gets0;$ \emph{//Max possible distance $\psi$ during $\Delta t$}\\
$\xi \gets 0;$ \emph{//Number of dimensions at $\Delta t$} \\
\For{$i, i=1, 2,\ldots,\xi$}{\emph{//Iterate number of dimensions during $\Delta t$}\\
    \If{$\exists WT_{P_{a}}[i](\Delta t) \vee \exists WT_{P_{b}}[i](\Delta t)$}{
        $\psi \gets \psi  + (\zeta * \zeta) $\;
        $\xi \gets \xi + 1 $\;
    }
}
$\psi \gets \sqrt{\psi}$\;
\emph{//Multi-dimensional distance computation}:\\
$\mu_{ab} \gets 0$\;
\For{$i, i=1, 2,\ldots,n$}{
    $x \gets 0;$ \emph{//Temporary variable}\;
    \uIf{$\exists WT^{s_i}_{P_{a}}(\Delta t)\wedge  \exists WT^{s_i}_{P_{b}}(\Delta t)$}{
        $x=\left|10^{WT^{s_i}_{P_{a}}(\Delta t).NRSSI}- 10^{WT^{s_i}_{P_{b}}(\Delta t).NRSSI}\right| $\;
    }
    \uElseIf{$\exists WT^{s_i}_{P_{a}}(\Delta t)$}{
        $x= 10^{WT^{s_i}_{P_{a}}(\Delta t).NRSSI}$\;
    }  
    \uElseIf{$\exists WT^{s_i}_{P_{b}}(\Delta t)$}{
        $x= 10^{WT^{s_i}_{P_{b}}(\Delta t).NRSSI}$\;
    }
    $\mu_{ab} \gets \mu_{ab} +x $\;
}
\uIf{$\sqrt{\mu_{ab}} \geq  \psi$}{
    \Return{$ \frac{\psi}{\xi}$\;}
}
\uElse{
    \Return{$ \frac{\sqrt{\mu_{ab}}}{\psi  * \xi}$\;}
}
\end{algorithm}

\begin{algorithm}
\footnotesize
\caption{Match score of wireless fingerprints} 
\label{Algo:FingerprintComp}
\SetKwInOut{Input}{Input}
\SetKwInOut{Output}{Output}
\Input{$WF_{P_{a}}(\Delta t)$,$WF_{P_{b}}(\Delta t)$}
\Input{$n$: Number of scanners during $\Delta T$}
\Input{$\upsilon$ (parameter): Match score vector}
\Output{$\delta_{ab}(\Delta t)$: Fingerprint-match score}
$\eta\gets0;$ \emph{//Max possible match score $\eta$ during $T$} \\
\For{$i, i=1, 2, ... n$}{
    $\eta \gets \eta + \upsilon[i]$\;
}
\emph{//Fingerprint match score computation}:\\
$\mu_{ab} \gets 0$\;
\For{$i, i=1, 2,\ldots, n$}{
    \If{$\exists WT_{P_a}[i](\Delta t)\wedge \exists WT_{P_b}[i](\Delta t)$}{
        \If{$WT_{P_a}[i](\Delta t).scanner$= $WT_{P_a}[i](\Delta t).scanner$}{
            $\mu_{ab} \gets \mu_{ab}  + \upsilon[i]$\;
        }
    }
}
    \Return{$\frac{\mu_{ab}}{\eta}$\;}
\end{algorithm}

\sloppy The second algorithm is based on creating matching scores of two wireless fingerprints in terms of their scanner orders, from higher $NRSSI$ to lower as previously shown in Eq.~\ref{eq:WF} and Eq.~\ref{eq:orderedlist2}. This algorithm has a parameter $\upsilon=(\upsilon[1], \upsilon[2],\ldots,\upsilon[n])$  (e.g., $\upsilon=[5,2,1]$ if $n=3$) which is called {\em match score vector}. This parameter weighs the matching scanner IDs according to their order in $WF$. We propose Algorithm~\ref{Algo:FingerprintComp} using $\upsilon$ for pairwise fingerprint match scores for time interval $T$ between $P_a$ and $P_b$. In this algorithm (for each $\Delta t$), the first iteration computes the maximum possible matching score based on the given $\upsilon$ and the number of scanners $n$ in the sampling time. The second iteration walks through the two ordered lists of $P_a$ and $P_b$ at the same time. If both have $WT$s in the given index $i$ and if both of the traces come from the same scanner, then it is regarded as a match and the corresponding match value $\upsilon[i]$ is given from the vector. For instance, if both $P_a$ and $P_b$ have $WT$s from all scanners $n$ and if their orders (from highest $NRSSI$ to lowest) are the same, the result of the second iteration is equal to the result of the first iteration (maximum possible match score). The ratio between the actual match score and the maximum possible match score gives the output $\delta_{ab}(\Delta t)$ that is the pairwise {\em fingerprint-match score} for the sampling time.

The aggregated {\em \textbf{W}ireless \textbf{F}ingerprint \textbf{M}atch} ($WFM$) (for the time interval $T$) between the pair is given by:
\begin{equation}
    \begin{centering}
    WFM_{ab}(T) = \frac{\sum _{\Delta t}^{\Theta_{ab}(T)} \delta_{ab}(\Delta t)}{\left |\Theta_{ab}(T)\right | } \text{ for } a \neq b. 
    \end{centering}
\end{equation}
$\Theta_{ab}(T)$ is the set of sampling times ($\Delta t \in T$) where both $P_a$ and $P_b$ have $WF$ (i.e., scanners observe both of them). $WFM$ does not aggregate the sampling times that only one device has $WF$ since the relative closeness of the unobserved person to the scanners is unknown.

\subsection{Decentralized Computing}
One of the main obstacles of group detection is the dimensional difference problem: How can we compare two $WF$s in the case of lacking scanners? 

In the RSSI trace comparison step of the decentralized computing, each scanner deals with its data independently. The main idea is that each scanner has its point of view; in other words, {\em perception}, which is possibly different from the view of another scanner. This perception of a scanner has a partial view of the world. Later, outputs of all scanners for the time interval $T$ are combined to create a {\em multi-scanner graph} and have a global view.

The decentralized scheme allows performing the computation on the scanner devices or nearby computation units. In this case, the task of each scanner is to perform the trace comparison locally. Thus, the problem of comparing RSSI data with different dimensions does not apply to this scheme as the wireless data for a scanner has only one dimension. For a scanner $s$, aggregated pairwise distance between every pair ($P_a$ and $P_b$) is called the {\em \textbf{S}ingle \textbf{P}erception \textbf{R}esult} ($SPR$), which is defined as follows:

\begin{equation}
  \begin{centering}
    d_{ab}(\Delta t)=\left|10^{WT^{s}_{P_{a}}(\Delta t).NRSSI}- 10^{WT^{s}_{P_{b}}(\Delta t).NRSSI}\right|,
  \end{centering}
\end{equation}
\begin{equation}
    \begin{centering}
    SPR_{ab}(T) = \frac{\sum _{\Delta t}^{\Theta_{ab}(T)} 
    d_{ab}(\Delta t)}{\left |\Theta_{ab}(T)\right | } \text{ for } a \neq b, 
    \end{centering}
\end{equation}
where $\Theta_{ab}(T)$ is the set of sampling times ($\Delta t \in T$) where both $P_a$ and $P_b$ have $WT$ in scanner $s$ (i.e., $s$ observes both of them).

$SPR$s of all observed pairs are encapsulated in a message and sent to the back-end server for finding the {\em \textbf{U}nified \textbf{P}erception \textbf{R}esult} ($UPR$), which is defined as follows:
\begin{equation}
    \begin{centering}
    UPR_{ab}(T) =  1- \frac{\sum _s^{S_{ab}(T)} SPR^{s}_{ab}(T)}{\left| S_{ab}(T) \right|  * \Omega }
    \text{ for } a \neq b,
    \end{centering}
\end{equation}
where $S_{ab}(T)$ is the set of scanners that have $SPR_{ab}$. $\Omega$ represents the maximum possible $SPR$ and it can be calculated considering maximum possible $NRSSI$ difference or set empirically. $UPR$s are used as the edge weights of the {\em multi-scanner graph}.

\begin{table}[]
\begin{tabular}{@{}lcc@{}}
\toprule
\textbf{Experiments}& \textbf{Controlled} &\textbf{Real-world} \\ \midrule
\# BT traces& $\sim$80K& $\sim$10M\\
\# people/beacons   & \{4, 7, 8\}& 14\\
\# BT scanners   &  3& 3\\
\# different rooms   & \{1, 2, 4\}& 4\\
Expected \# of groups  & \{1, 2, 3, 4\}      & 6\\
Advertising interval       & [0.1,1]sec     & 0.1sec\\
Transmission power         & [-4, -12]dBm     & -4dBm\\
\# of different setups & 27    & 1\\
Data duration & $\sim$5 hours& $\sim$25 days \\ 
BT data storage & $\sim$23.5MB & $\sim$2.2GB\\
\# parameter trials-WFM   &  10 & 1 \\
\# param. trials-Group detection  &  10 & 1 \\ \bottomrule       
\end{tabular}
\caption{Information related to the experiments.}
\label{Table:Experiments}
\end{table}

\begin{table*}
  \caption{The algorithms used in the experiments and the ranges for parameter values. The algorithms and their places as part of the approach are in Fig.~\ref{Fig:Approach} (except the Girvan-Newman, DBScan, or MeanShift, which are used for comparison).}
  \label{tab:calibrationparameters}
  \begin{tabular}{lllc}
  \toprule
  \textbf{Analytics algorithms}& \textbf{Approach phases}& \textbf{Parameters}&\textbf{Parameter trial values}\\  \midrule
DenGraph& Group detection& Cluster distance&$(0,1{]}$\\
HCS& Group detection& Min edge weight - Threshold&$(0,1{]}, (0, 1.5{]}$\\
MaxClique& Group detection&Min edge weight - Threshold & $(0,1{]}, (0, 1.5{]}$\\
WFM& Fingerprint trajectory match& $\upsilon$: Match score vector& ${[}{[}3,100{]}, {[}2, 25{]}, {[} 1, 5{]}{]}$  \\
MDD & Fingerprint trajectory match& $\zeta$: Max distance parameter & 7 (empirical) \\ \hline
\textbf{Benchmark algorithms}\\ \hline 
Girvan-Newman&Graph clustering (after WFM)  &-&-\\ 
DBScan& Clustering preprocessed data& $\epsilon$: Epsilon&   (0,1{]}\\ 
MeanShift&Clustering preprocessed data &Kernel   &   (0,1{]}  \\

 \bottomrule

\end{tabular}
\end{table*}

\subsection{Group Detection}
We define a graph $G= (V, E, w)$, where $V=\{v_1, v_2,\ldots,v_n\}$ is the set of vertices and $n$ is the number of people observed by the scanners during time interval $T$, $E\subseteq V x V$ is the set of edges, and $w:E\mapsto \mathbb{R}_+$ is the weight function. As a vertex of the graph represents the device of a person, the vertex representing $P_a$ is denoted as $v_a$ and $w_{ab}$ denotes edge weight with $v_b$. The edge weight values can be given by $MDD$, $WFM$, or $UPR$. Larger weight values indicate matching fingerprints.

Checking the edge weight between two people is not sufficient to decide if they belong to a group since the group characteristics differ from the pairwise relations. As a simple example, considering three people $P_a, P_b, P_c$, the values $w_{ab}$ and $w_{bc}$ can be large while $w_{ac}$ is small. The subject of group characteristics is studied extensively, considering various measures such as intra- and inter-group distances and betweenness centrality. 

For the group detection step, we apply three graph clustering algorithms that are also in past studies related to social networks. The first is called density-based community detection algorithm (DenGraph~\cite{falkowski2007dengraph}), which is a modification of the well-known density-based spatial clustering of applications with noise (DBScan~\cite{ester1996density}) algorithm. The main difference of DenGraph is that it performs clustering on graph models. The other two clustering techniques used are Highly Connected Subgraphs (HCS~\cite{hartuv2000clustering}) and Maximal Cliques (MaxClique~\cite{makino2004new}). These two clustering algorithms are similar to each other in the sense that they both generate connected subgraphs. HCS is a less constrained version of MaxClique where the enumerated subgraphs do not need to be fully connected. The clustering algorithms run in different time intervals and output cluster labels for each vertex. Each label represents a group, and we classify vertices with the same labels as a group.

\subsection{Long-Term Linkages}
\label{Long-Term}
While the groups dynamically change over time, Group-In saves the group information for every time interval in the {\em group database}. The long-term linkage step involves several statistical group analysis. First, Group-In analyzes the closeness of two people based on the number of time intervals ($T$s) that they are listed in the same group in the database and the number of time intervals where both of them are present. Based on the ratio between the two values, it estimates a long-term linkage value. Having the linkage values between every pair results in the long-term linkage graph. The resulting graph for long periods can be given as an input for the analysis of the social relationship networks by social scientists, which may lead to the development of new ways to improve interactions between people in areas such as a campus or an office environment. For instance, in an office environment, the face to face interactions of different teams can be compared, and further assistance can be given to more socially isolated people. Group-In long-term linkage statistics can be used as one of the parameters for analyzing these relationships. Moreover, Group-In creates long-term group statistics. These statistics involve the number of detected people, the number of individuals vs. groups, distribution of group sizes, and the ratio between the number of detected people and inferred groups.

%% file: Experiment.tex
\begin{figure}
\centering
    \begin{subfigure}[b]{0.22\textwidth}
\includegraphics[width=0.95\columnwidth]{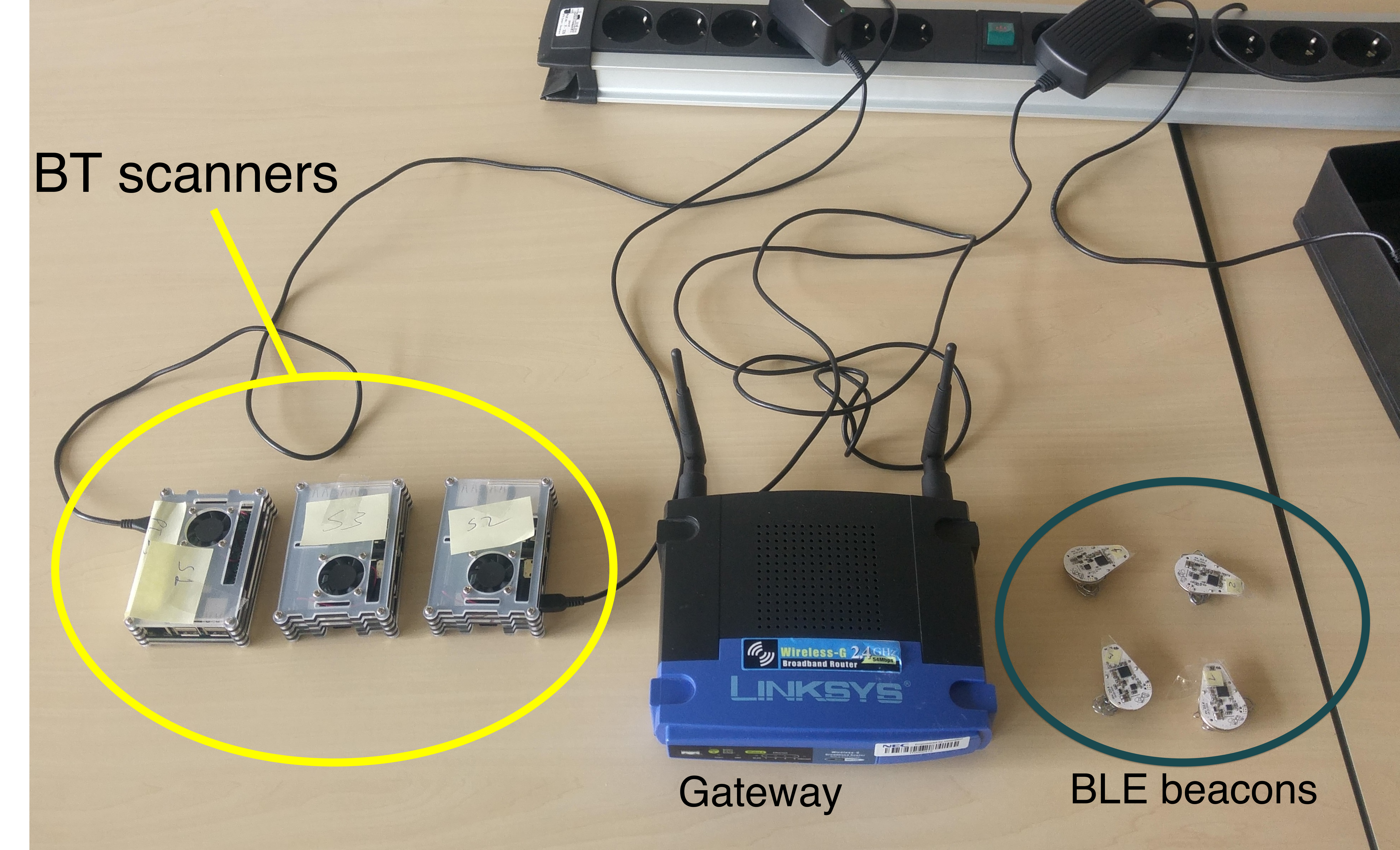}
\caption{Some devices used in the experiments.}
\label{Fig:ControlledDevices}
\end{subfigure}~
    \begin{subfigure}[b]{0.235\textwidth}
\includegraphics[width=1.05\textwidth]{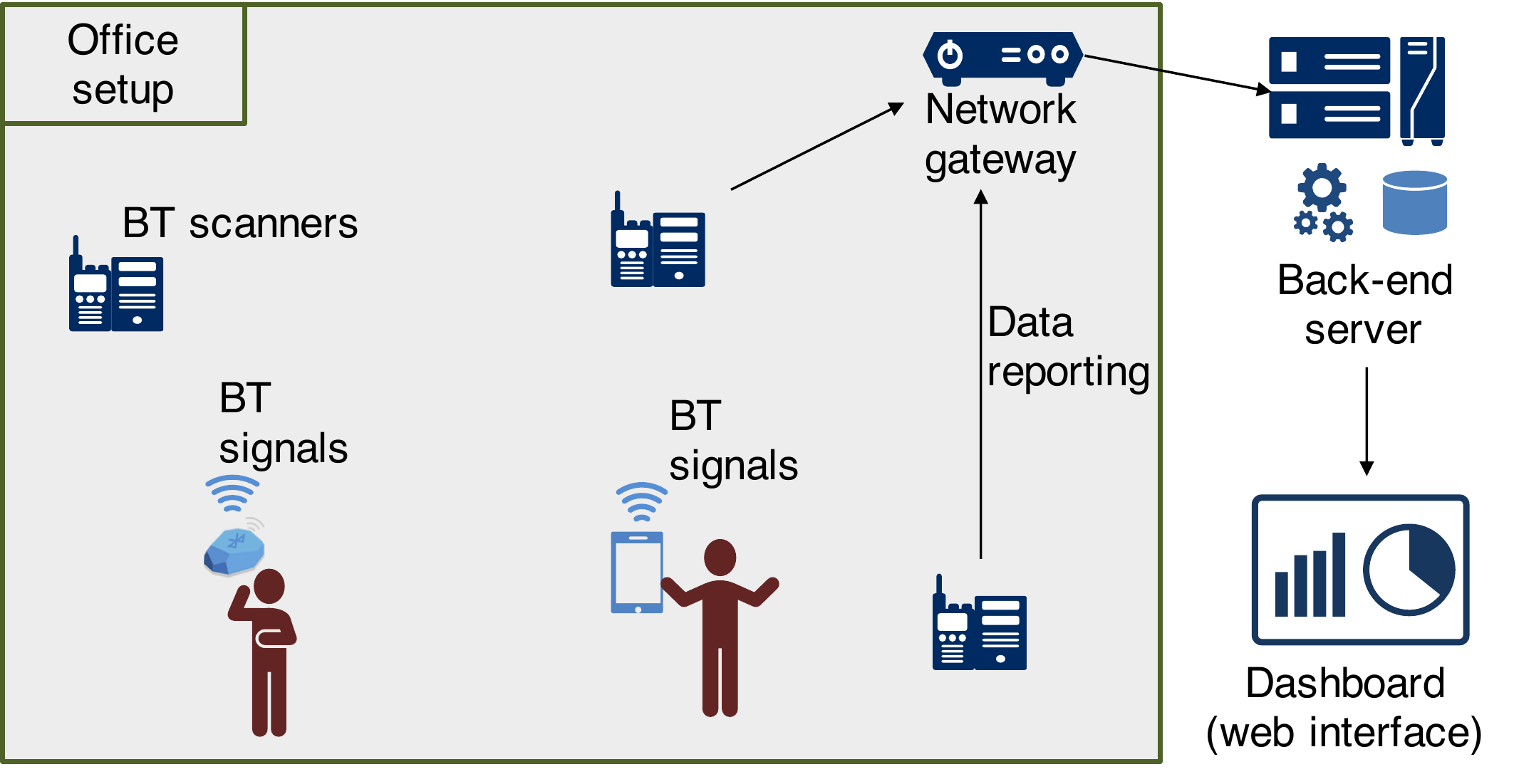}
\caption{Illustration of the Group-In system setup.} \label{Fig:SystemBasic}
\end{subfigure}
\caption{Experimental setup: Devices and placement.}
\end{figure}

\section{Experimental Evaluation}
\label{Evaluation}
\subsection{System Setup}

Fig.~\ref{Fig:SystemBasic} illustrates the basic system setup of Group-In. The setup consists of three wireless scanners deployed in an area with the capability to receive packets from people's mobile devices. Distances between the scanners are  $\sim$10~m. The wireless scanners can perform computation on the devices or send their raw measurements to a back-end server through the network gateway. The back-end server has data brokering and analytics modules as well as storage capability using a NoSQL database (CouchDB), where the wireless traces are indexed based on their timestamps. Group-In visualizes the offline or real-time results coming from the server on the web dashboard.

\subsection{Experimental Settings}

\begin{figure}
    \centering
    \begin{subfigure}[b]{0.22\textwidth}
        \includegraphics[width=1.01\textwidth]{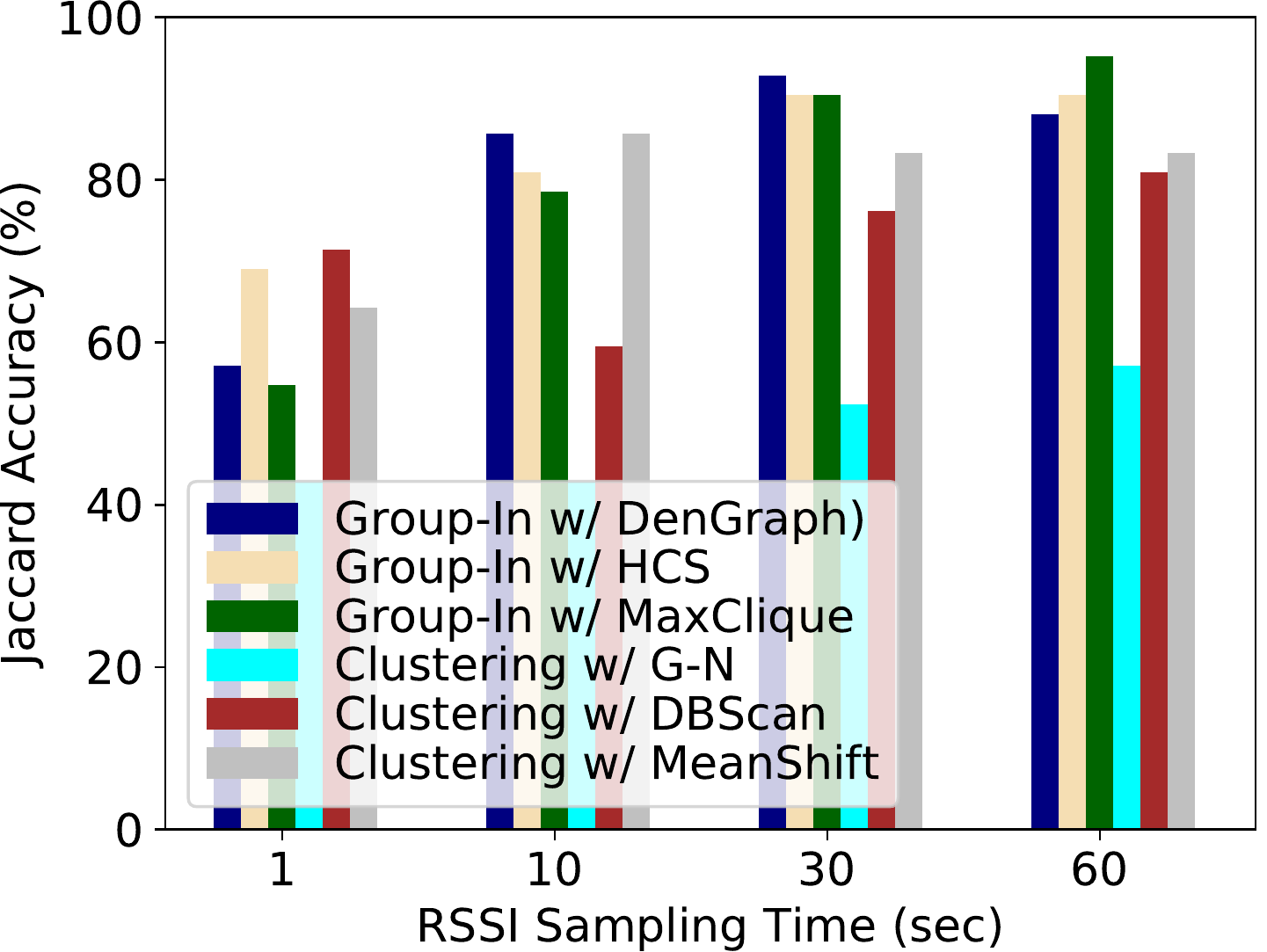}
    \end{subfigure}
    ~
    \begin{subfigure}[b]{0.22\textwidth}
        \includegraphics[width=1.01\textwidth]{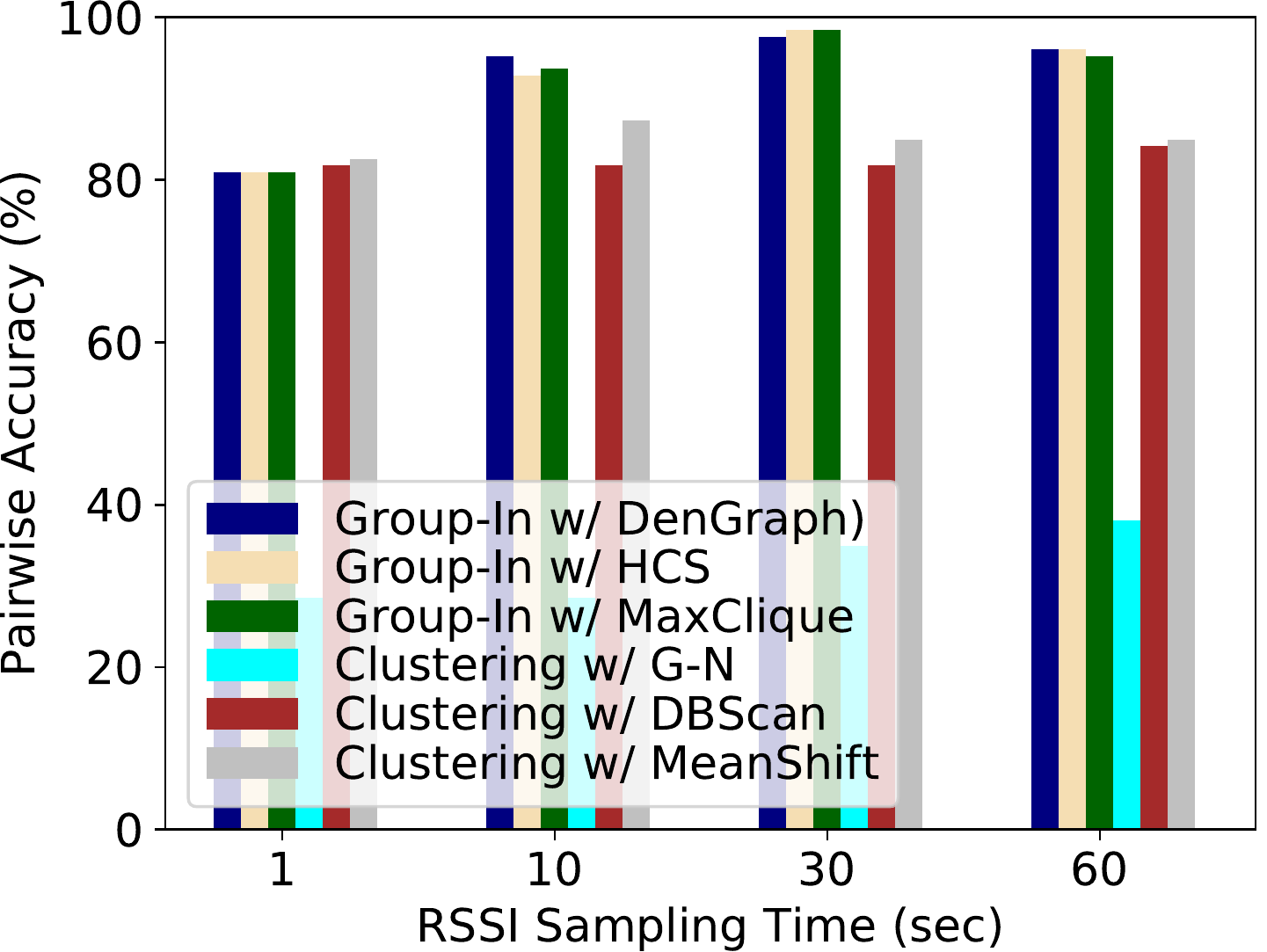}
    \end{subfigure}
    
    \begin{subfigure}[b]{0.22\textwidth}
        \includegraphics[width=1.01\textwidth]{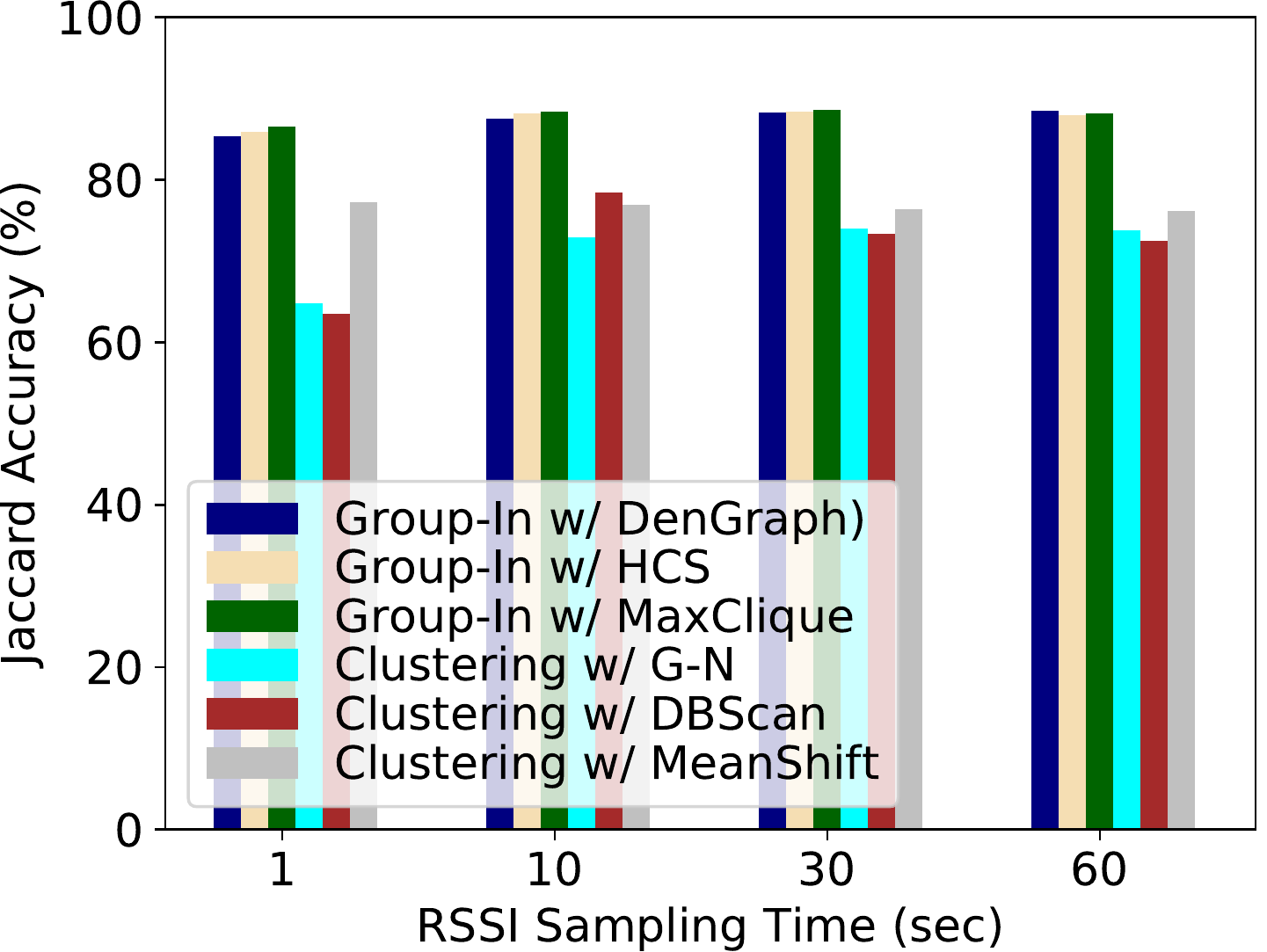}
                                \vspace*{-3mm}
    \end{subfigure}
    ~
    \begin{subfigure}[b]{0.22\textwidth}
        \includegraphics[width=1.01\textwidth]{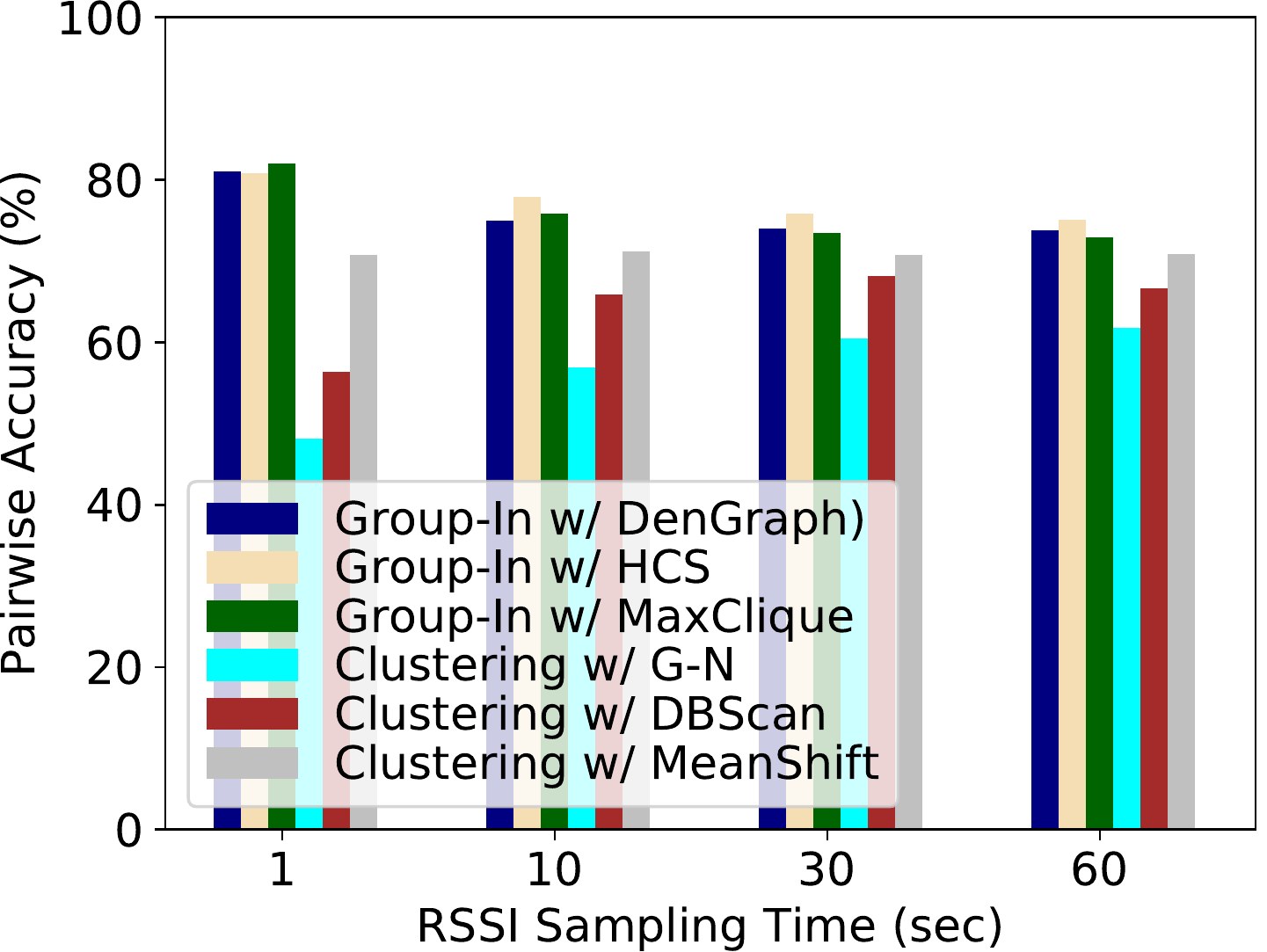}
        \vspace*{-3mm}
    \end{subfigure}
    \caption{The results of the office scenario (centralized). Top: controlled, bottom: real-world.}\label{Exp1-cent}
\end{figure}

We conduct two experimental studies for testing the group inference accuracy in various settings. The first is a set of controlled lab experiments, with devices shown in Fig.~\ref{Fig:ControlledDevices}. We conduct the controlled experiments through short data collection campaigns (each has about 10~min duration). The second is a real-world experiment in an office environment, with 14 employees for more than one month. The controlled experiments consist of 27 different settings. These settings include having beacons placed in 1, 2, or 4 rooms, where 4, 7, or 8 beacons are distributed. The parameter values of these experiments (summarized in Table~\ref{Table:Experiments}) aim to cover different scenarios including detecting people in different rooms, groups in the same room, detecting static/mobile groups, differentiating movement trajectories (e.g., Group 1 and 2 in Fig.~\ref{Fig:GroupingGoal}), and detecting groups with gradually shorter distances to each other (from 10 to 1~m). In the second (real-world) experiment, 12 people work in four rooms in the vicinity of the scanners, and two people are visitors from distant rooms. The employees stick BLE beacons to their access badges that they carry for a month. We labeled the controlled and real-world datasets with ground truth information where the groups of the devices are known based on static/mobile placements or working places of the employees.

\begin{figure}
    \centering
    \begin{subfigure}[b]{0.22\textwidth}
        \includegraphics[width=1.01\textwidth]{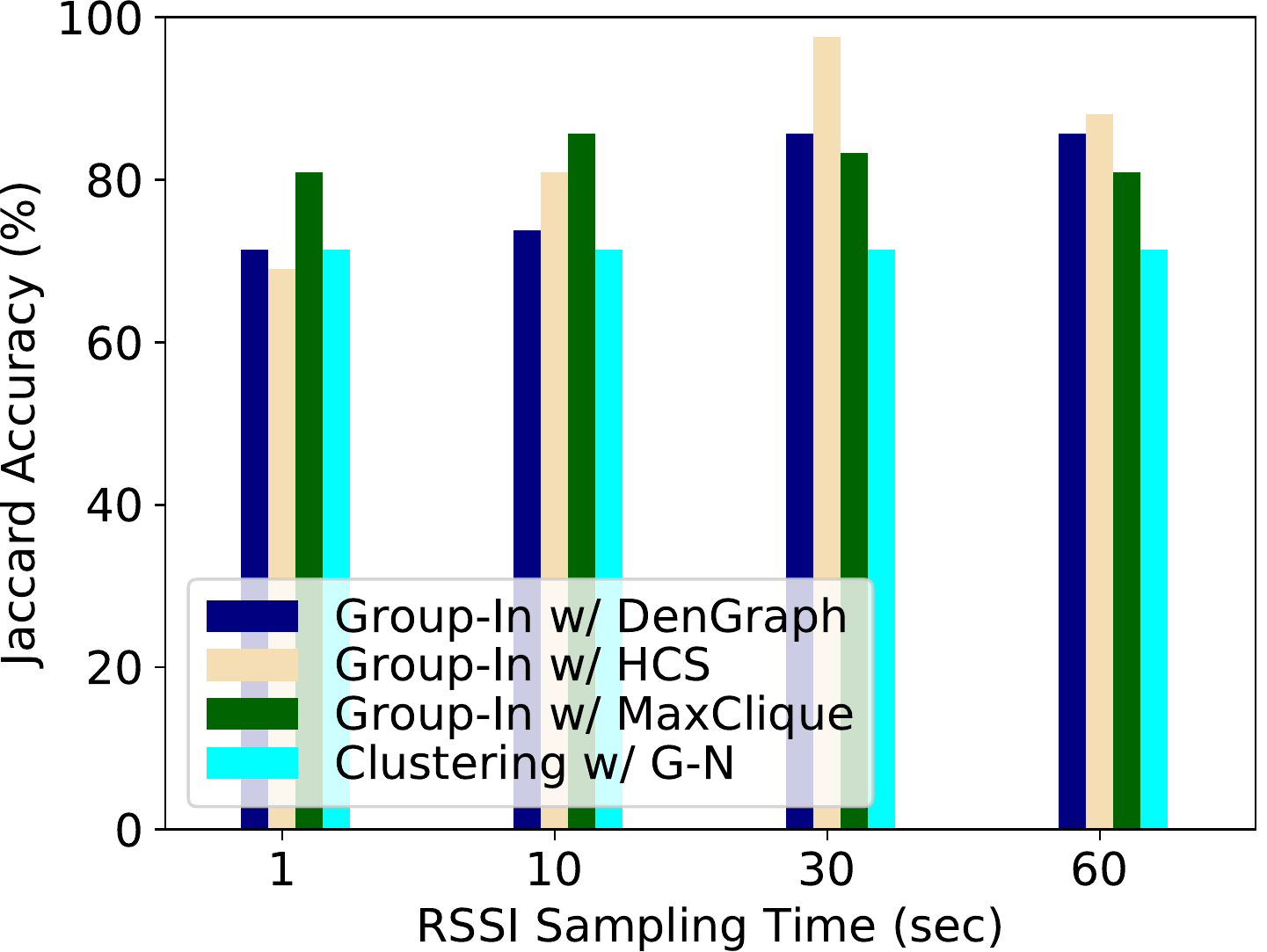}
    \end{subfigure}
    ~
    \begin{subfigure}[b]{0.22\textwidth}
        \includegraphics[width=1.01\textwidth]{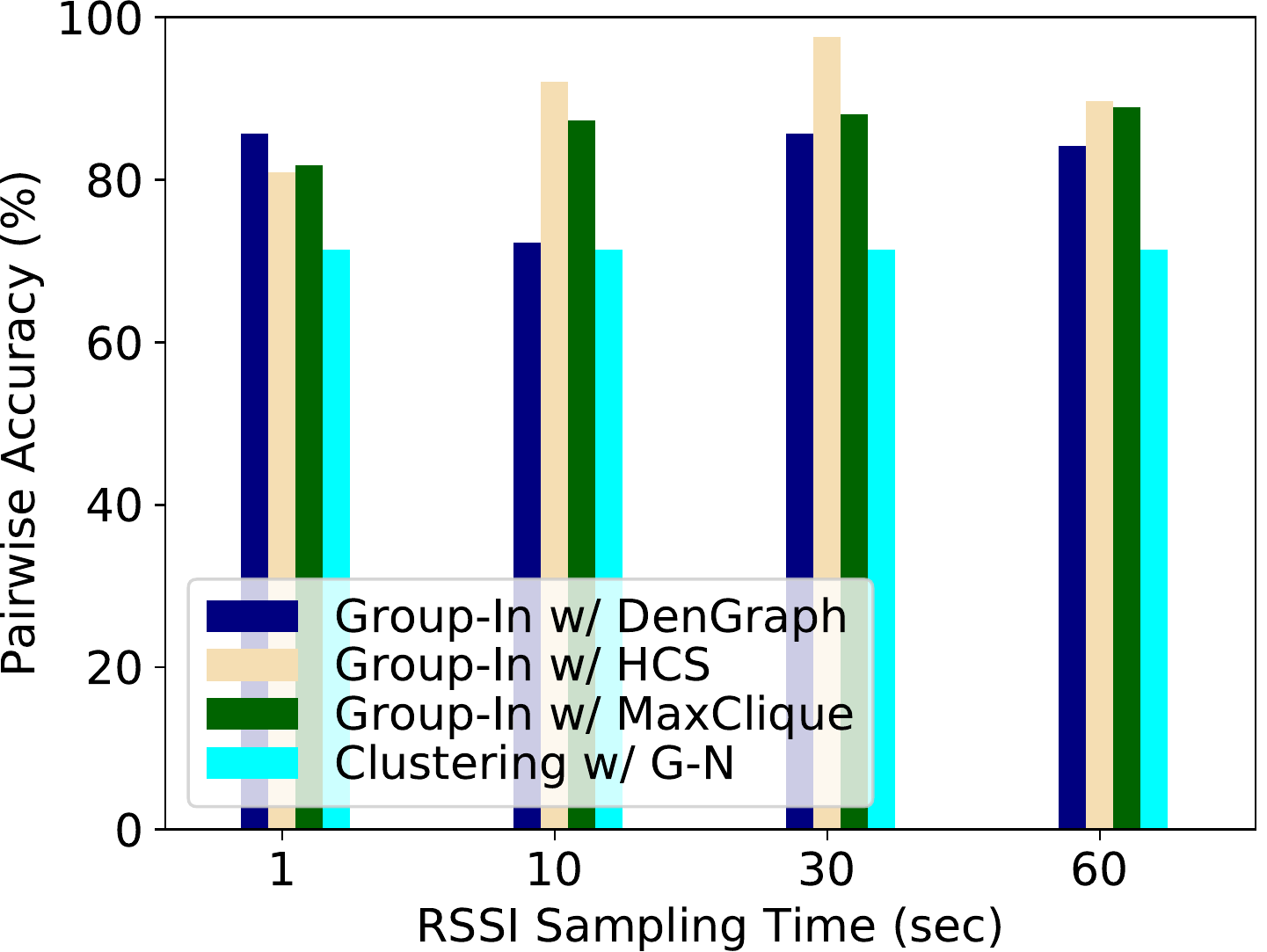}
    \end{subfigure}
    
    \begin{subfigure}[b]{0.22\textwidth}
        \includegraphics[width=1.01\textwidth]{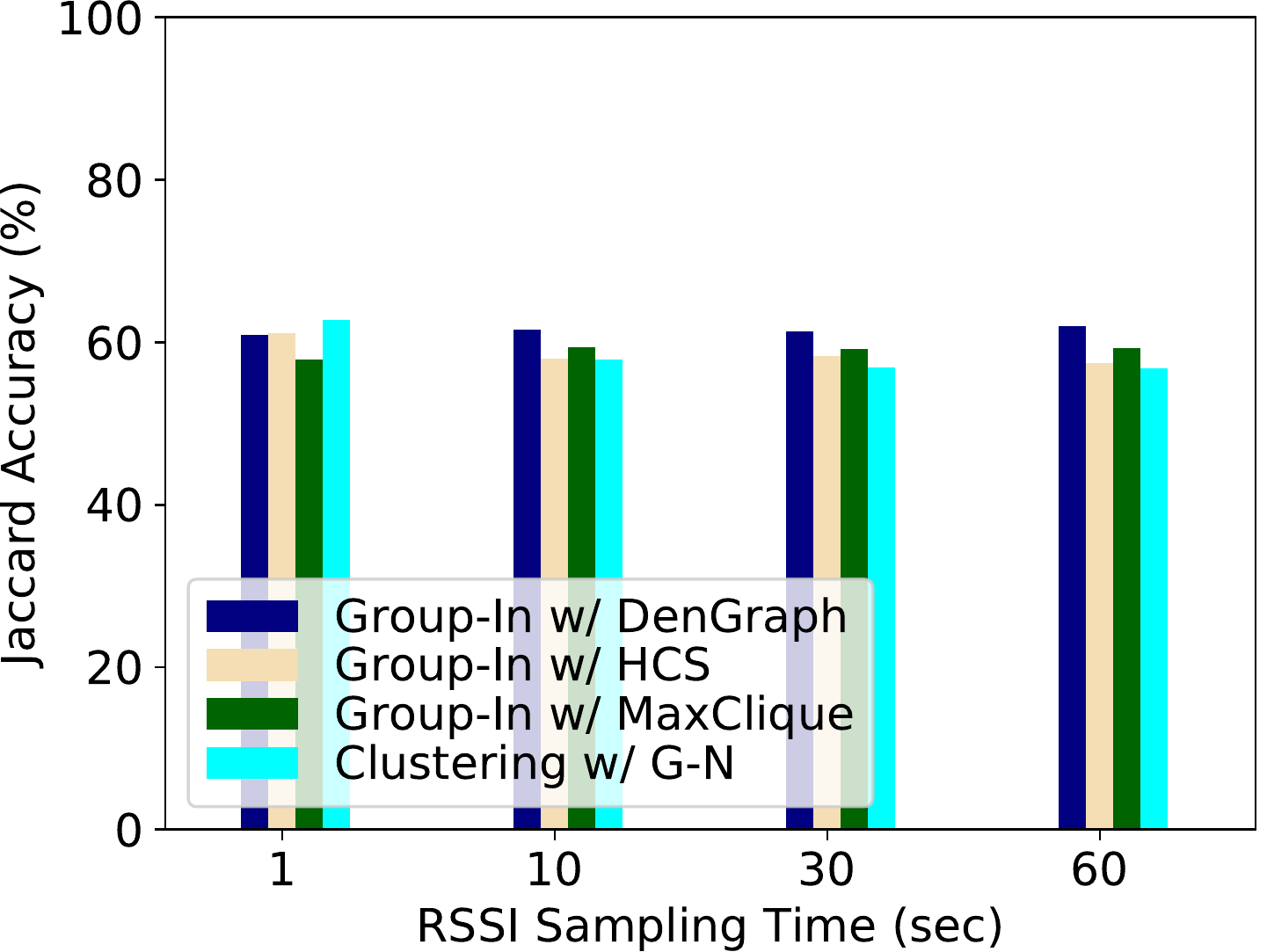}
    \end{subfigure}
    ~
    \begin{subfigure}[b]{0.22\textwidth}
        \includegraphics[width=1.01\textwidth]{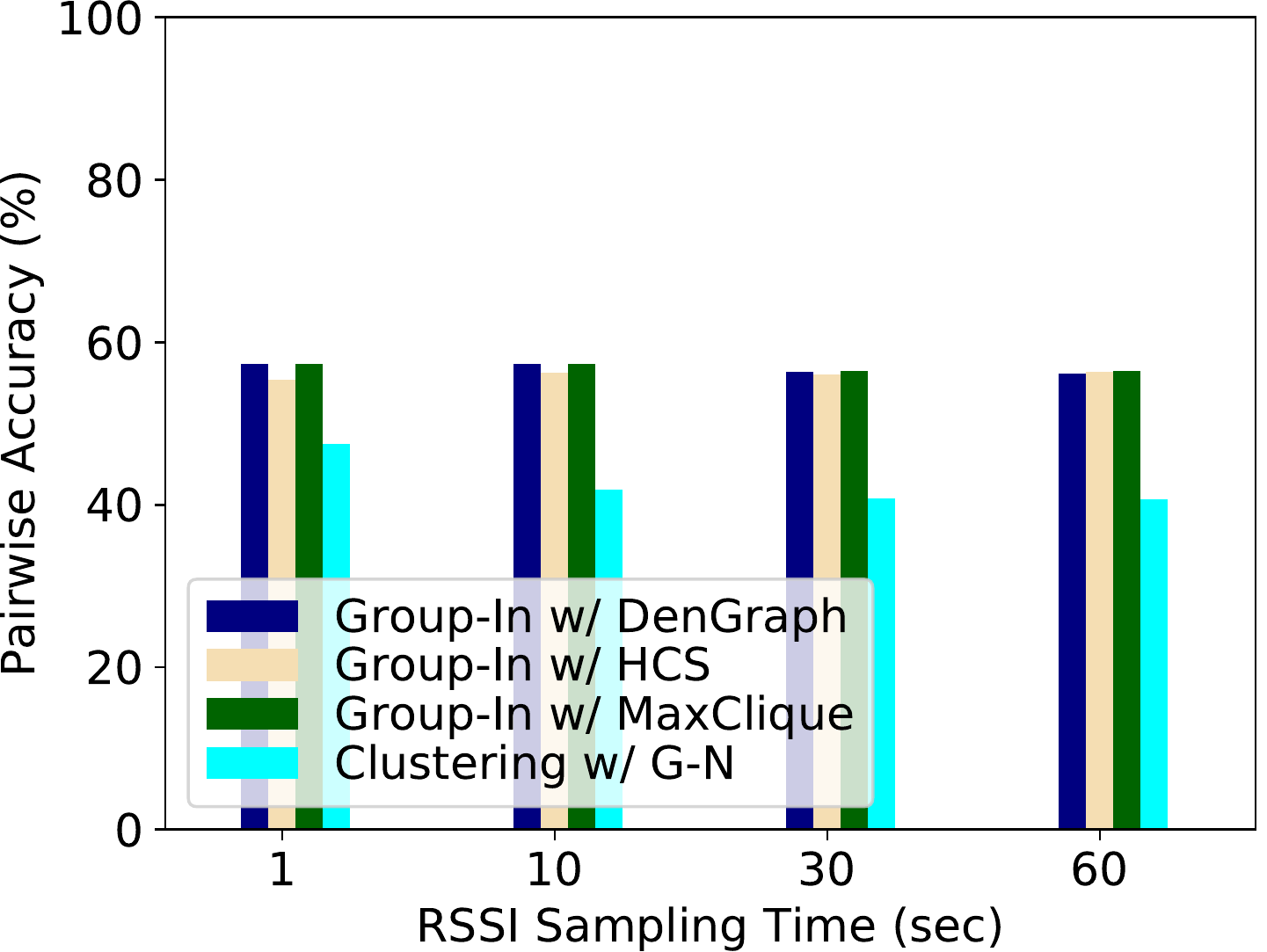}
    \end{subfigure}
    
    \caption{The results of the office scenario (decentralized). Top: controlled, bottom: real-world.}\label{Exp1-dist}
\end{figure}

Table~\ref{tab:calibrationparameters} lists the algorithms that we use in the analysis. We test the proposed approach using DenGraph, HCS, MaxClique for group detection step, and WFM and MDD for trajectory matching step (previously shown in Fig.~\ref{Fig:Approach}). The approach is compared against applications of the popular clustering algorithms DBScan~\cite{ester1996density} and MeanShift ~\cite{Comaniciu:2002:MSR:513073.513076} directly after the preprocessing steps. Furthermore, we test Girvan-Newman ~\cite{Girvan2002Community} (shown as 'G-N' in figures) for comparison as it is a community detection algorithm popularly used for graph data and social network analysis. Girvan-Newman (G-N) algorithm uses the output graph from the fingerprint distance aggregation step using WFM. Each algorithm calibrates itself by stochastic trials of 10 parameter values inside the given ranges (parameter trial values in Table~~\ref{tab:calibrationparameters}). The results in the controlled experiments are cross-validated using 20\% of the training data. We apply the learned parameter values from the controlled experiments to the real-world setup without any training. Unless otherwise stated, the default values in the experiments are $\Delta t=5$~sec and $T=120$~sec using WFM and HCS for the centralized computing and UPR and HCS for the decentralized computing.

\begin{figure}
    \centering
    \begin{subfigure}[b]{0.22\textwidth}
        \includegraphics[width=1.01\textwidth]{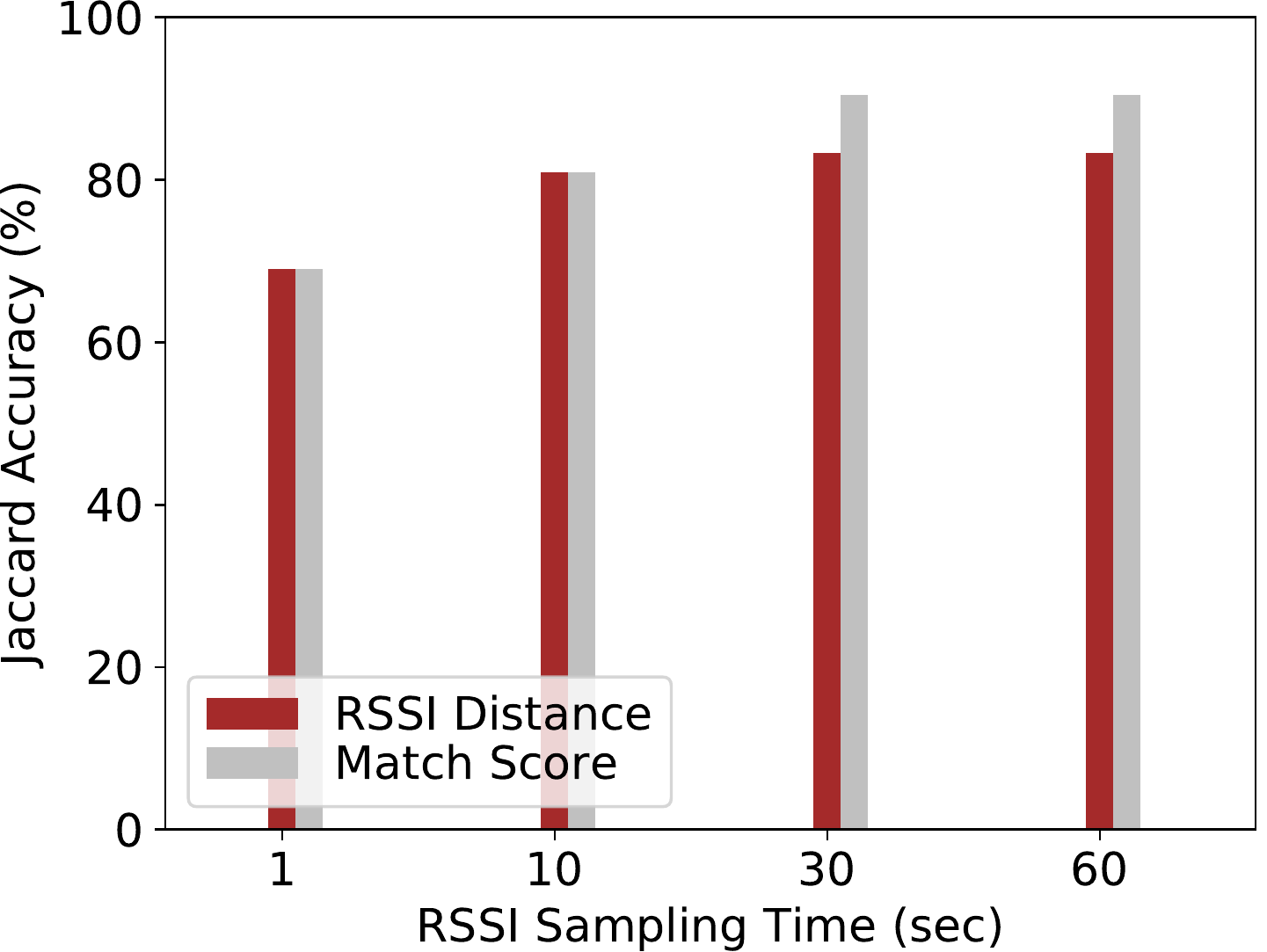}
        \vspace*{-3mm}
    \end{subfigure}
    ~
    \begin{subfigure}[b]{0.22\textwidth}
        \includegraphics[width=1.01\textwidth]{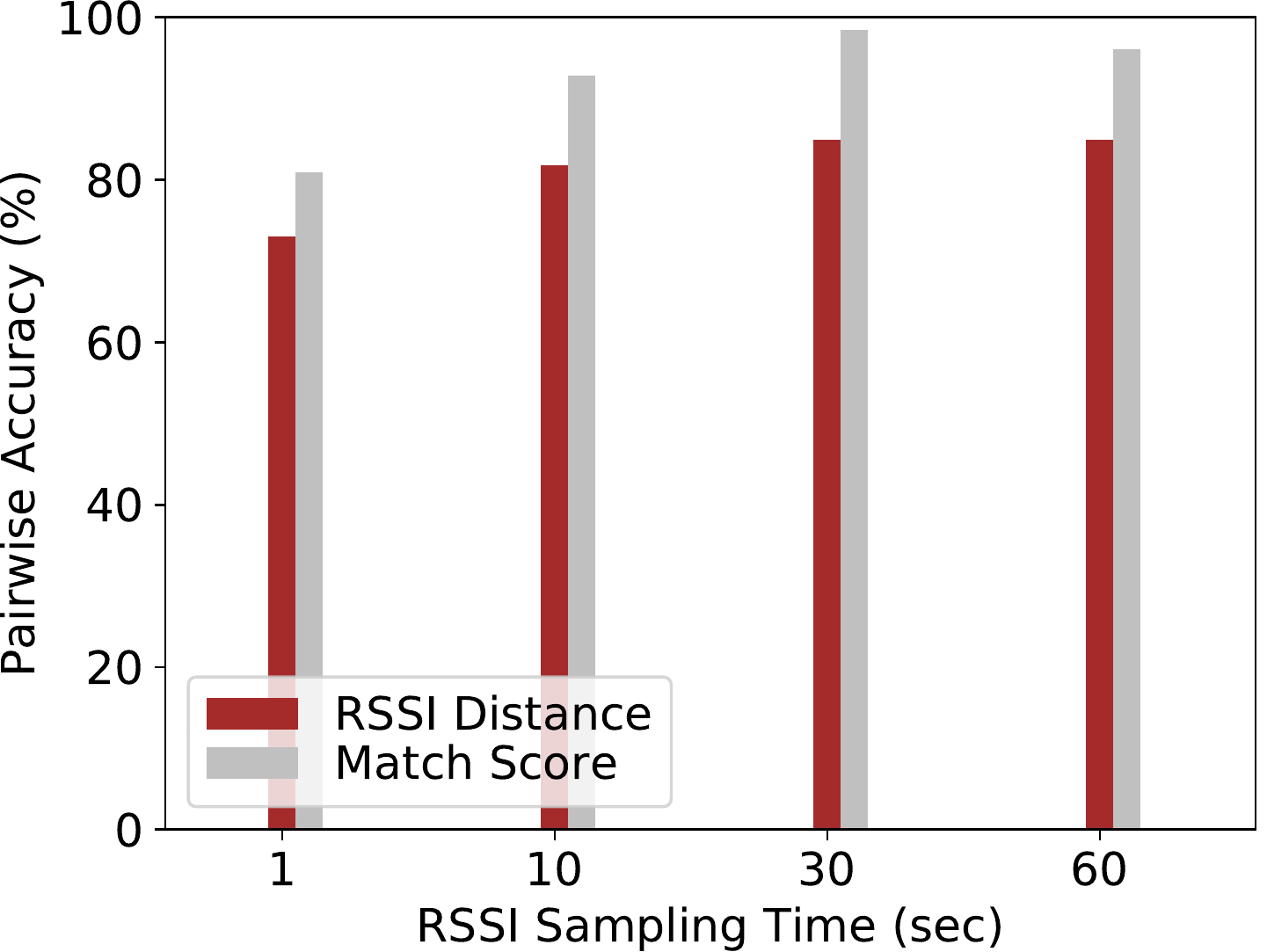}
                        \vspace*{-3mm}
    \end{subfigure}
    
    \caption{The results of the fingerprint trajectory match algorithms for the office scenario (controlled).}\label{Exp1-RSSI}
\end{figure}

We use two accuracy metrics: 1) {\em Pairwise similarity coefficient} and 2) {\em The Jaccard index}. The pairwise coefficient is a result of pairwise comparisons for all pairs present at $T$. It is calculated as $\dfrac{TP+TN}{C(|P|,2)}$ where $C(|P|,2)$ is the pair combinations among the people $P$. $TP$ (true positives) is the number of pairs in the same groups who are classified with the same labels. $TN$ (true negatives) is the number of pairs in different groups who are classified with different labels. The Jaccard index is given by $J(A,B) =\frac{|A \cap B|}{|A \cup B|}, |A \cup B| = |P| $. For the Jaccard index, we first recursively match every observed group with the ground truth group which has the largest intersection (most number of shared members). Then, for every person, we check if the person is classified with the correct label (the ground truth label) to find the size of the intersection set $|A \cap B|$. As a benchmark one can consider the random placement of people into groups. Considering having 4 groups in the controlled scenarios and 6 groups in the real-world scenario, a random guess may result in about 16-25\% accuracy in most of the cases for both of the metrics.

\begin{figure}
    \centering
    \begin{subfigure}[b]{0.22\textwidth}
        \includegraphics[width=1.01\textwidth]{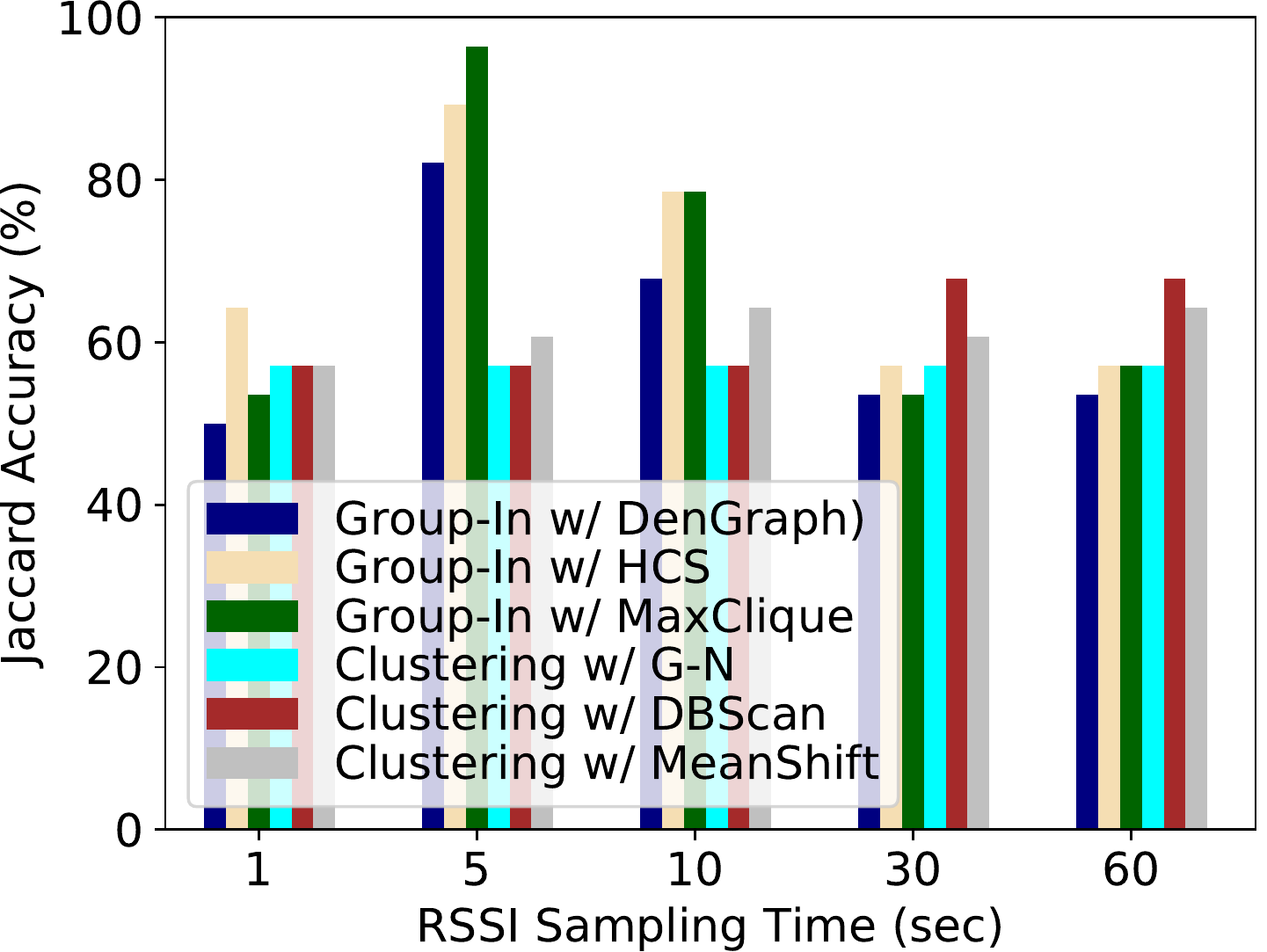}
    \end{subfigure}
    ~ 
    \begin{subfigure}[b]{0.22\textwidth}
        \includegraphics[width=1.01\textwidth]{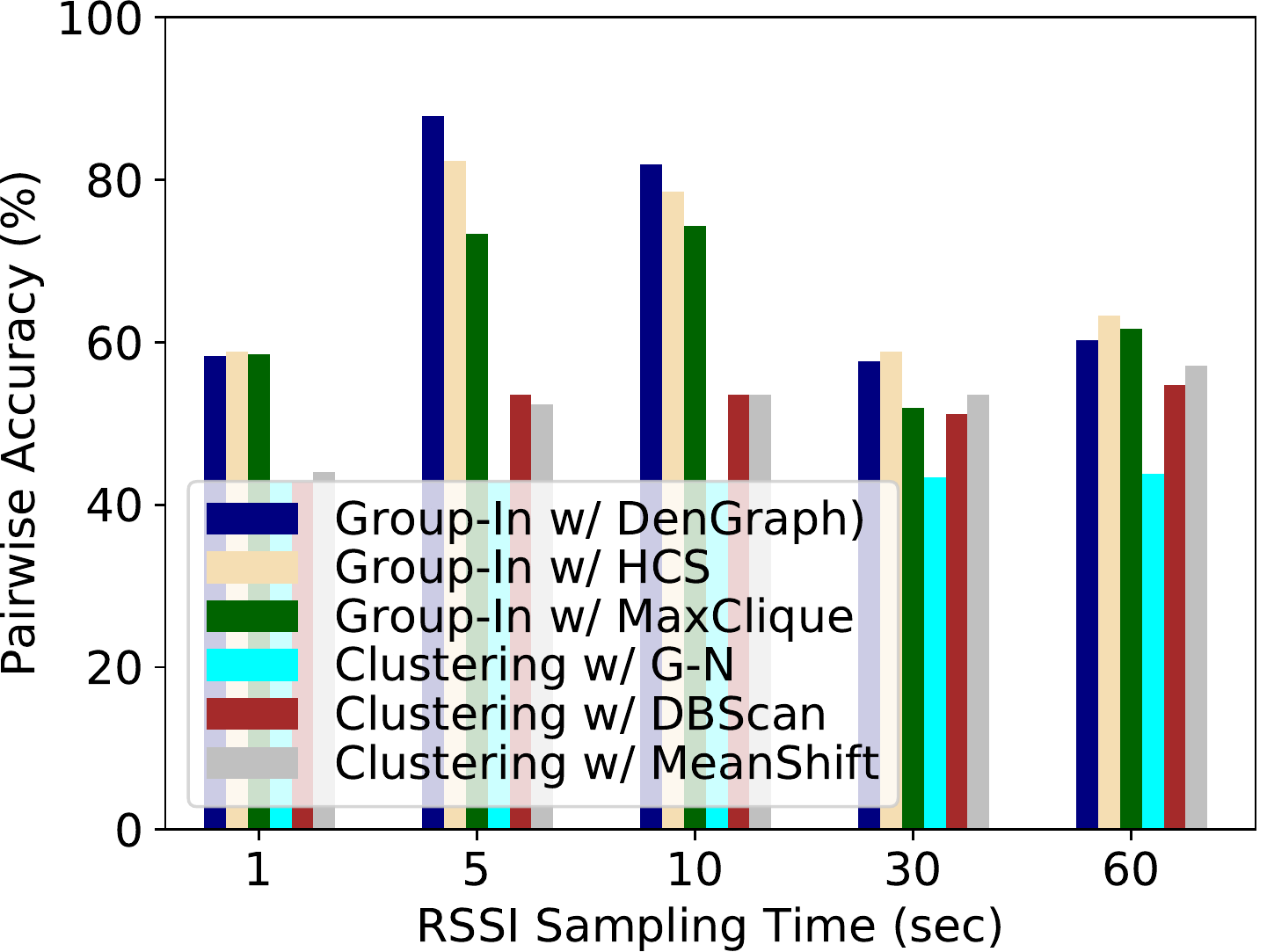}
    \end{subfigure}
  
    \begin{subfigure}[b]{0.22\textwidth}
        \includegraphics[width=1.01\textwidth]{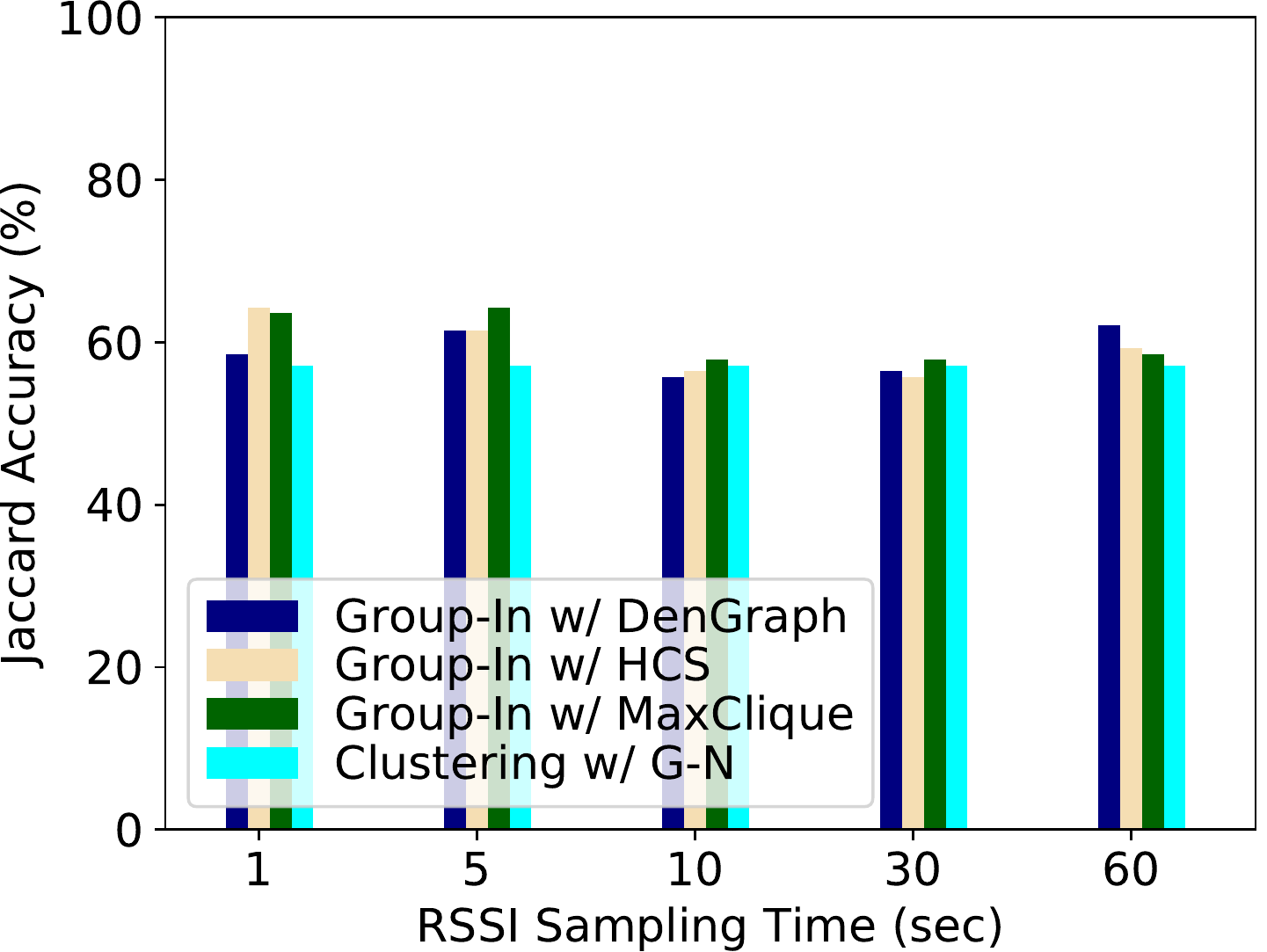}
                                \vspace*{-3mm}
        
    \end{subfigure}
    ~ 
    \begin{subfigure}[b]{0.22\textwidth}
        \includegraphics[width=1.01\textwidth]{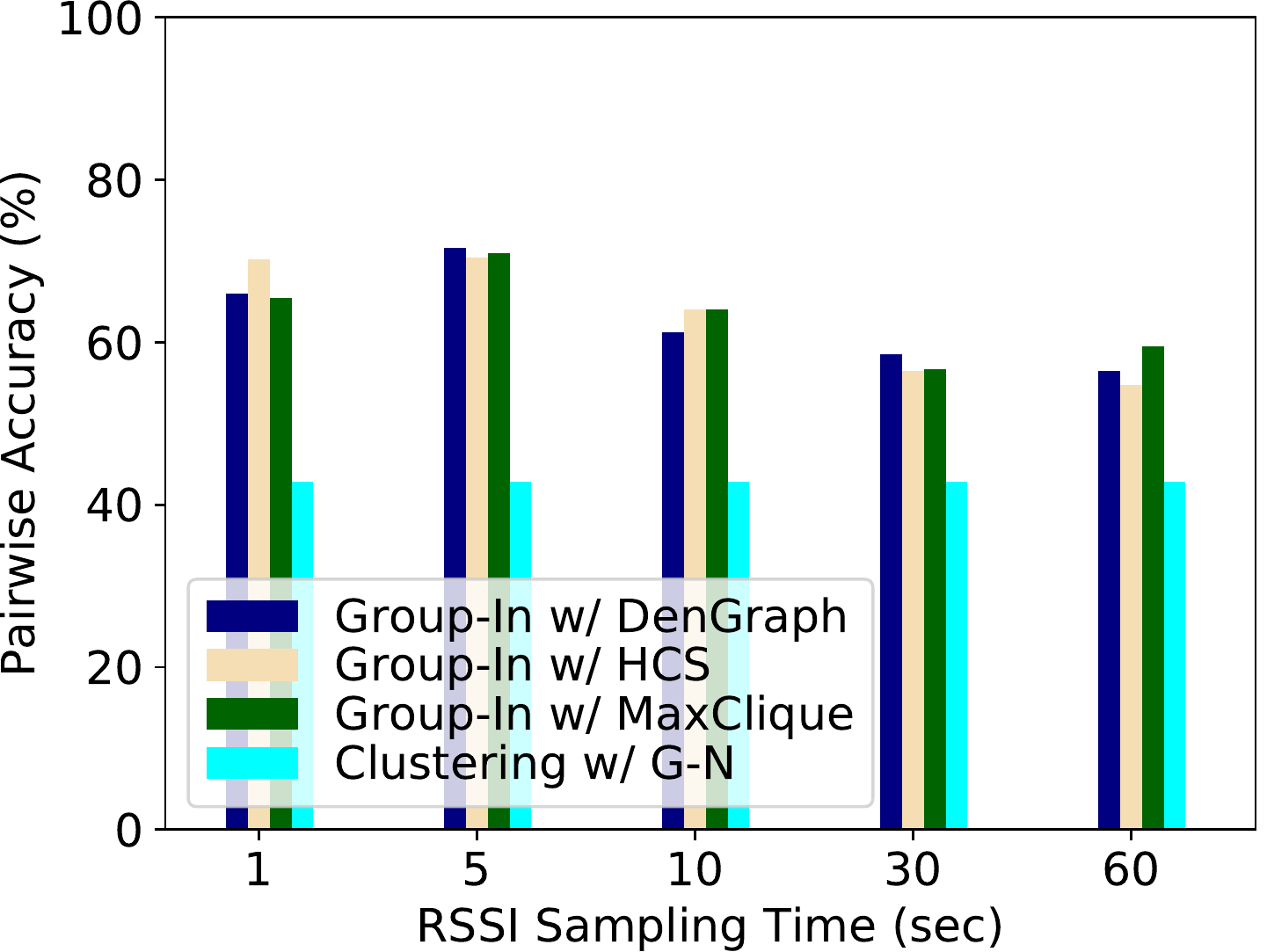}
                                \vspace*{-3mm}
        
    \end{subfigure}
    \caption{The results of the straight-walking scenario. Top: centralized, bottom: decentralized.}\label{fig:MobileExpStraight}
\end{figure}

\subsection{Experimental Results}

This subsection starts with the experimental results for the office scenario where we analyze Group-In for office setups and the mobile scenario where we simulate different movements of groups. We also analyze the effects of the distance between two groups by placing the groups gradually closer in the controlled environment. Furthermore, we include the real-time and offline monitoring interface and long-term analysis results from our real-world experiment. Lastly, we include our remarks and discuss the limitations we observed during the experiments.

\noindent
\textbf{Results from the office scenario:} The first set of results we include from the controlled and real-world experiments. The first one simulates an office environment, whereas the second uses data collected from employees during their daily work schedules. The scanners' positions are the same for both experiments. There exist four office rooms and two corridors in between, where one scanner is in a room, and two scanners are on the corridors. For the controlled setup, seven beacons (mimicking seven people) are distributed into up to four groups and statically placed to the rooms. The results in Fig.~\ref{Exp1-cent}, Fig.~\ref{Exp1-dist} and Fig.~\ref{Exp1-RSSI} compare pairwise (left) and Jaccard (right) accuracy of the centralized and decentralized computing w.r.t. different sampling times.

Fig.~\ref{Exp1-cent} shows the results of centralized computing. DenGraph, HCS, and MaxClique achieve more than 90\% accuracy for controlled experiments with 30 or 60~sec sampling times. On the other hand, DBScan and MaxClique provide relatively higher accuracy compared to Girvan-Newman. Moreover, results demonstrate that graph clustering algorithms that use WFM (HCS, MaxClique, DenGraph) have higher accuracy than directly applying clustering algorithms (DBScan and MeanShift) in both the controlled and real-world setups.

\begin{figure}
    \centering
    \begin{subfigure}[b]{0.22\textwidth}
        \includegraphics[width=1.01\textwidth]{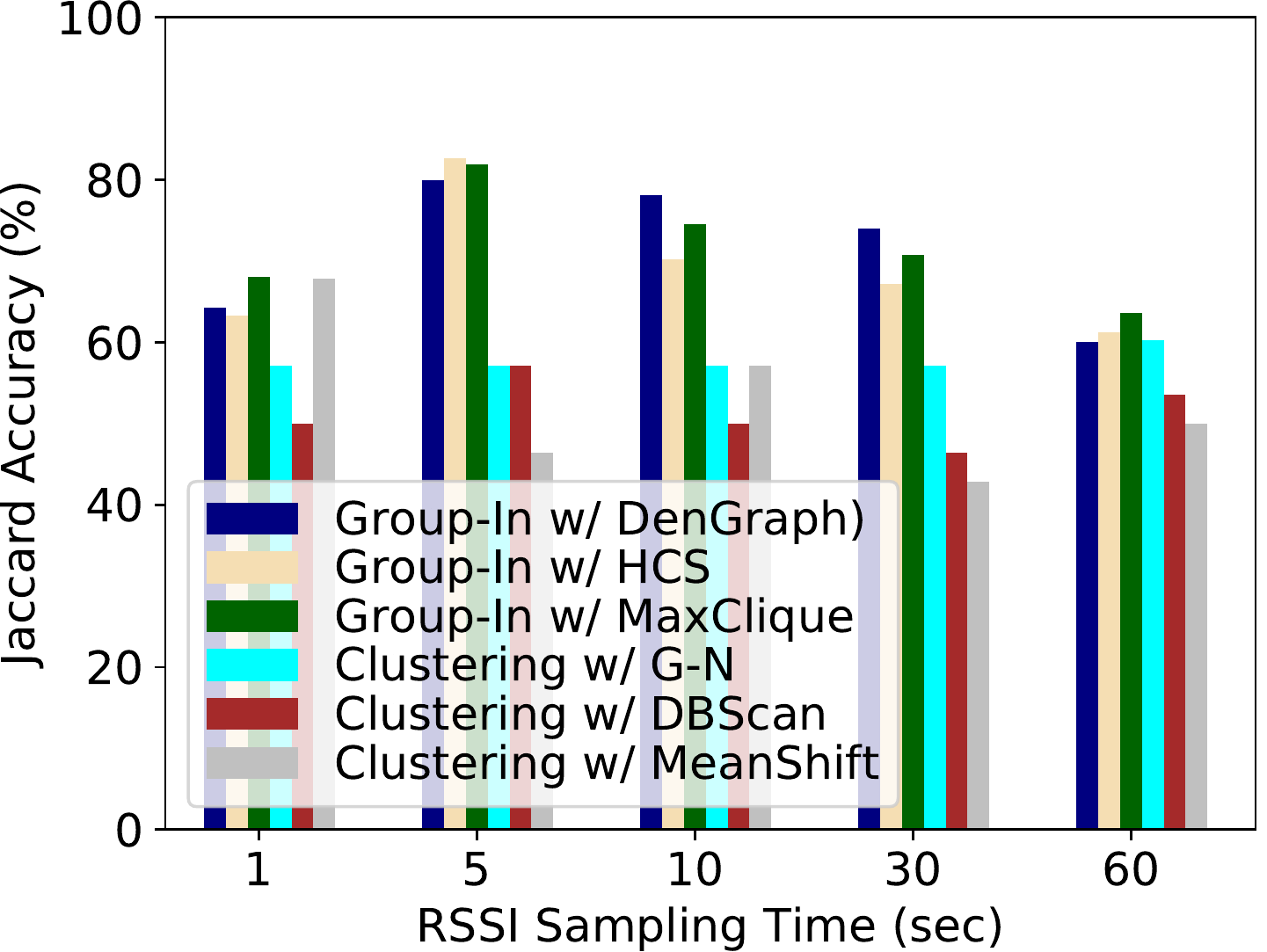}
    \end{subfigure}
    ~ 
    \begin{subfigure}[b]{0.22\textwidth}
        \includegraphics[width=1.01\textwidth]{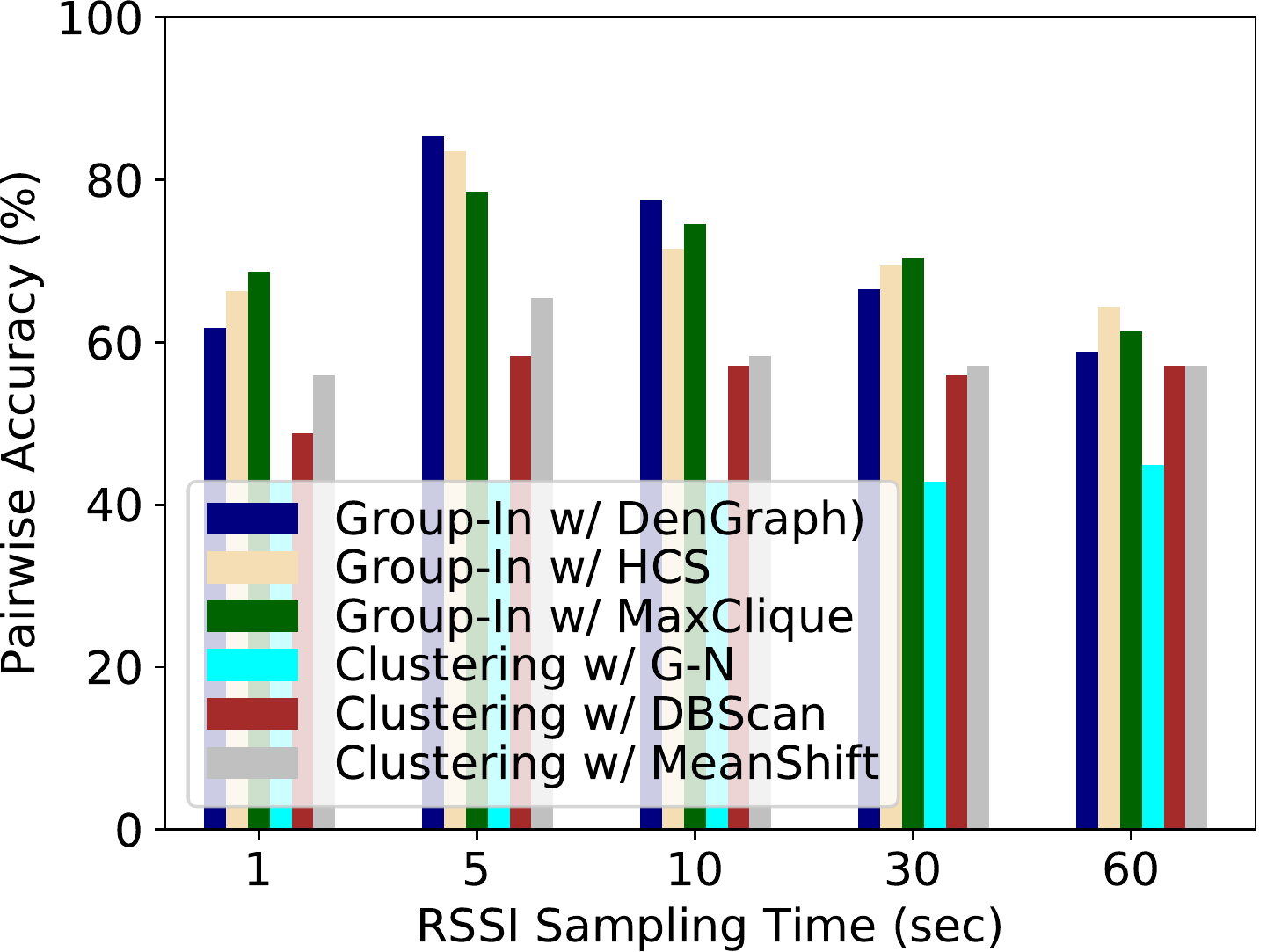}
    \end{subfigure}

    \begin{subfigure}[b]{0.22\textwidth}
        \includegraphics[width=1.01\textwidth]{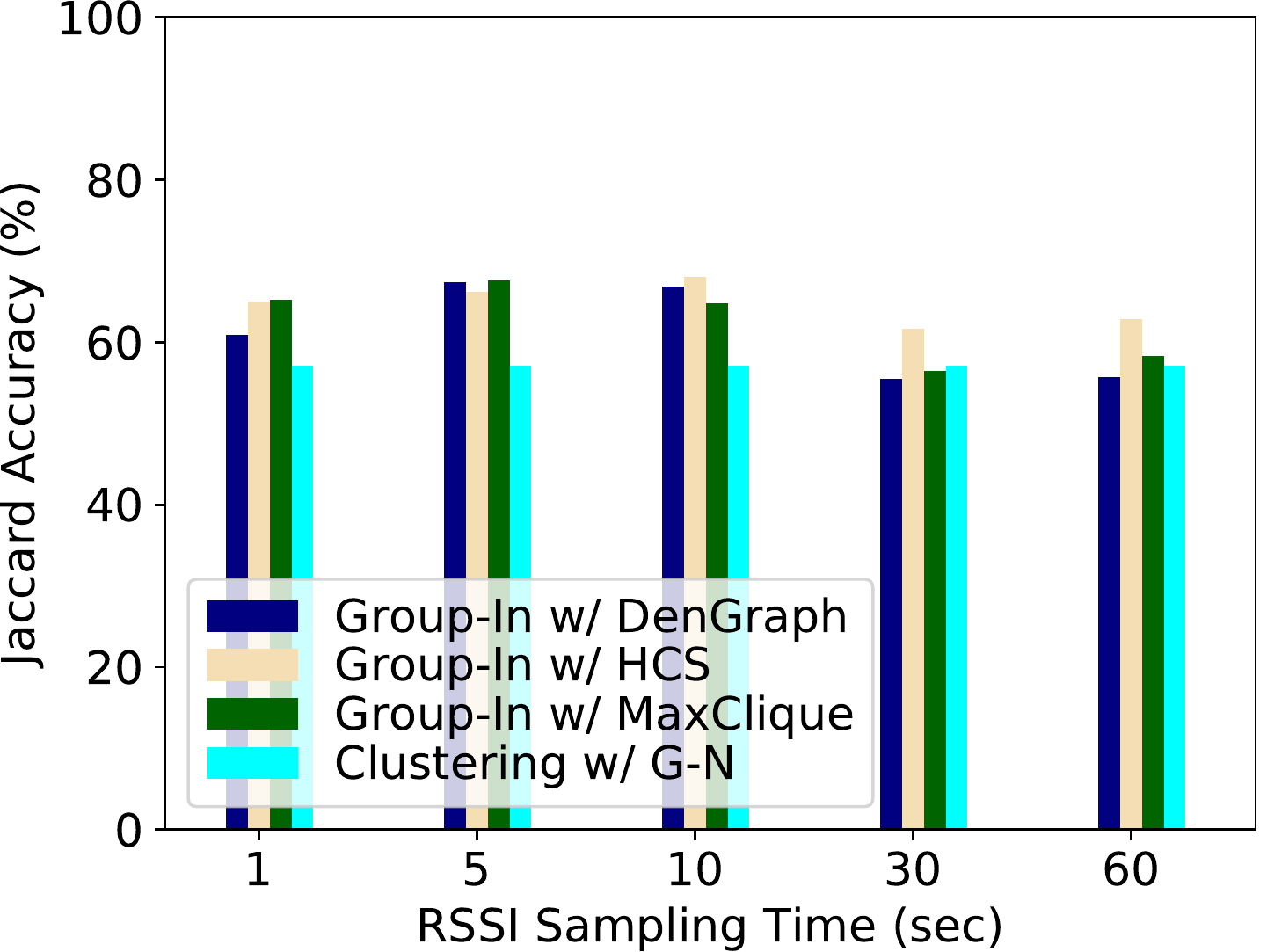}
                                        \vspace*{-3mm}
        
    \end{subfigure}
    ~ 
    \begin{subfigure}[b]{0.22\textwidth}
        \includegraphics[width=1.01\textwidth]{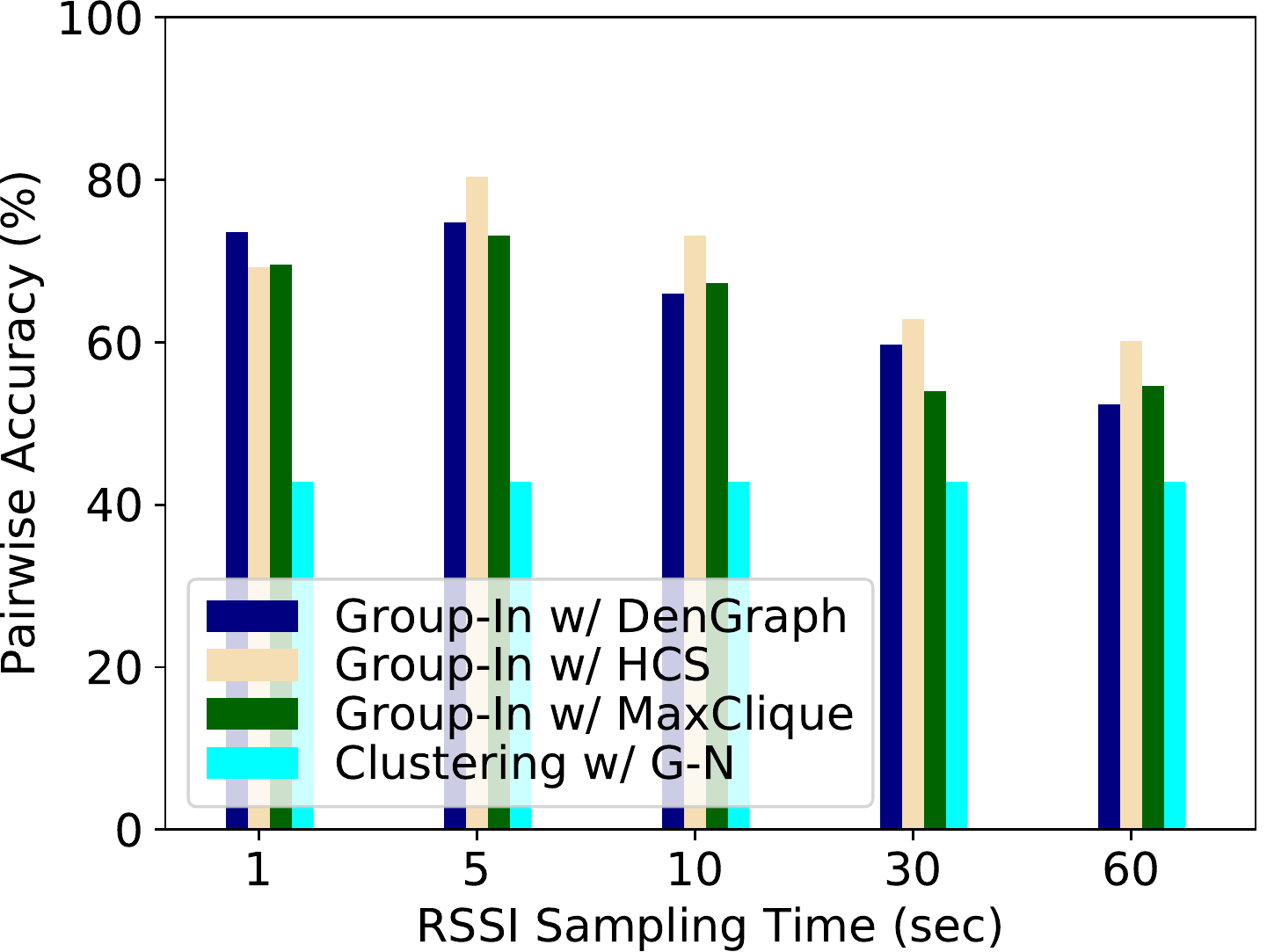}
                                        \vspace*{-3mm}
        
    \end{subfigure}
    \caption{The results of the random-walk scenario. Top: centralized, bottom: decentralized. }\label{fig:MobileExpRandom}
\end{figure}

Fig.~\ref{Exp1-dist} presents the accuracy results of decentralized computing. Decentralized computing achieves high accuracy in controlled experiments. For example, HCS achieves 98\% accuracy with $\Delta t=30$~sec. Besides, the accuracy of the decentralized computing in the real-world has a low accuracy (about 60\%) where group detection algorithms perform similarly, and Girvan-Newman has even lower pairwise accuracy. The sparse deployment of the scanners can cause this inaccuracy. In the real-world setup, some scanners classify people in two separate rooms in the same group if both of these rooms are distant from the scanner. Fig.~\ref{Exp1-RSSI} presents the results of using match scores (WFM) and multi-dimensional distances (MDD) (both with HCS). Both approaches produce high accuracy, especially with the increased sampling times, whereas the WFM approach achieves slightly higher accuracy.

\begin{figure}
    \centering
    \begin{subfigure}[b]{0.22\textwidth}
            \includegraphics[width=1.01\textwidth]{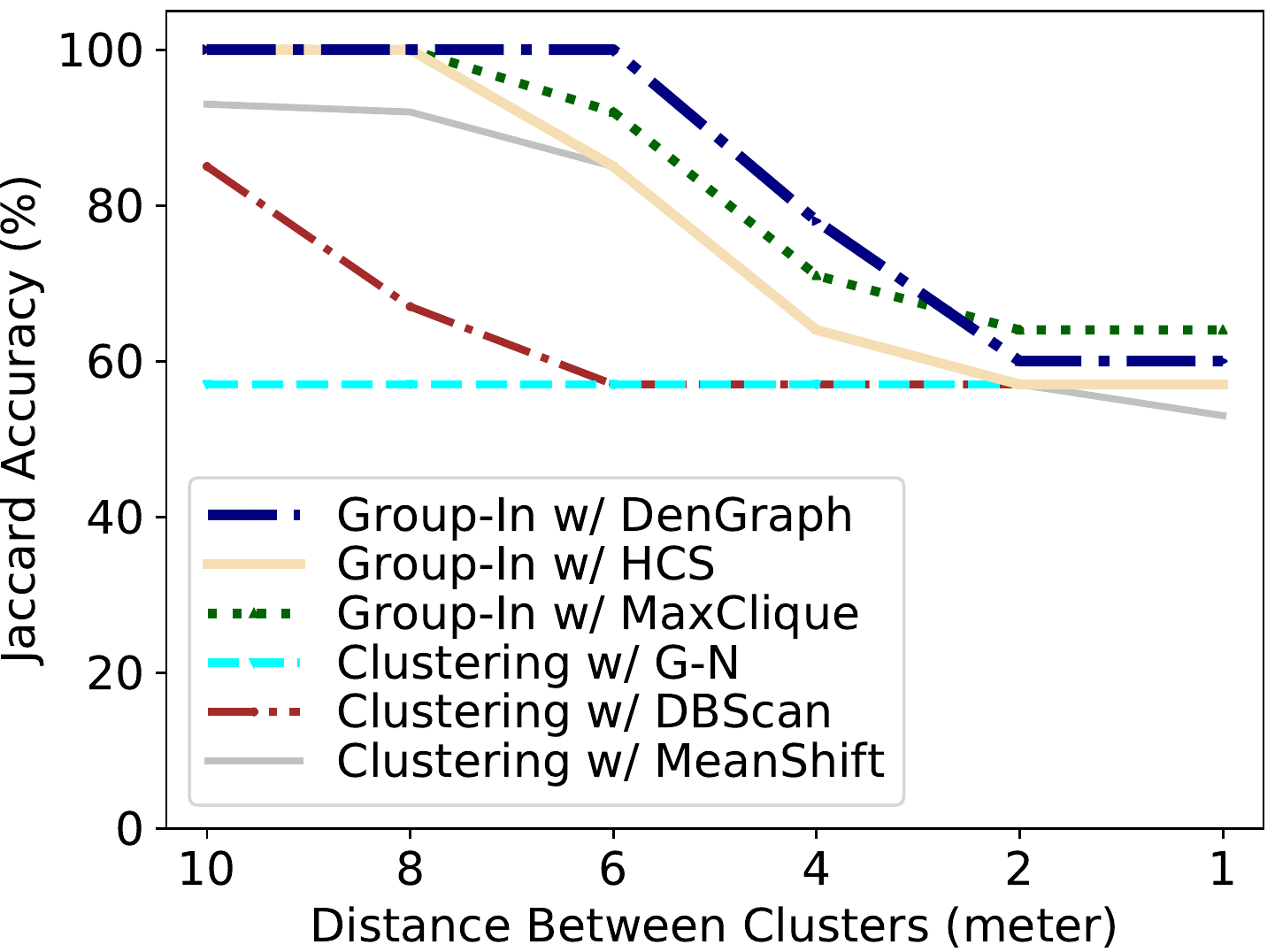}
    \end{subfigure}
    ~
    \begin{subfigure}[b]{0.22\textwidth}
        \includegraphics[width=1.01\textwidth]{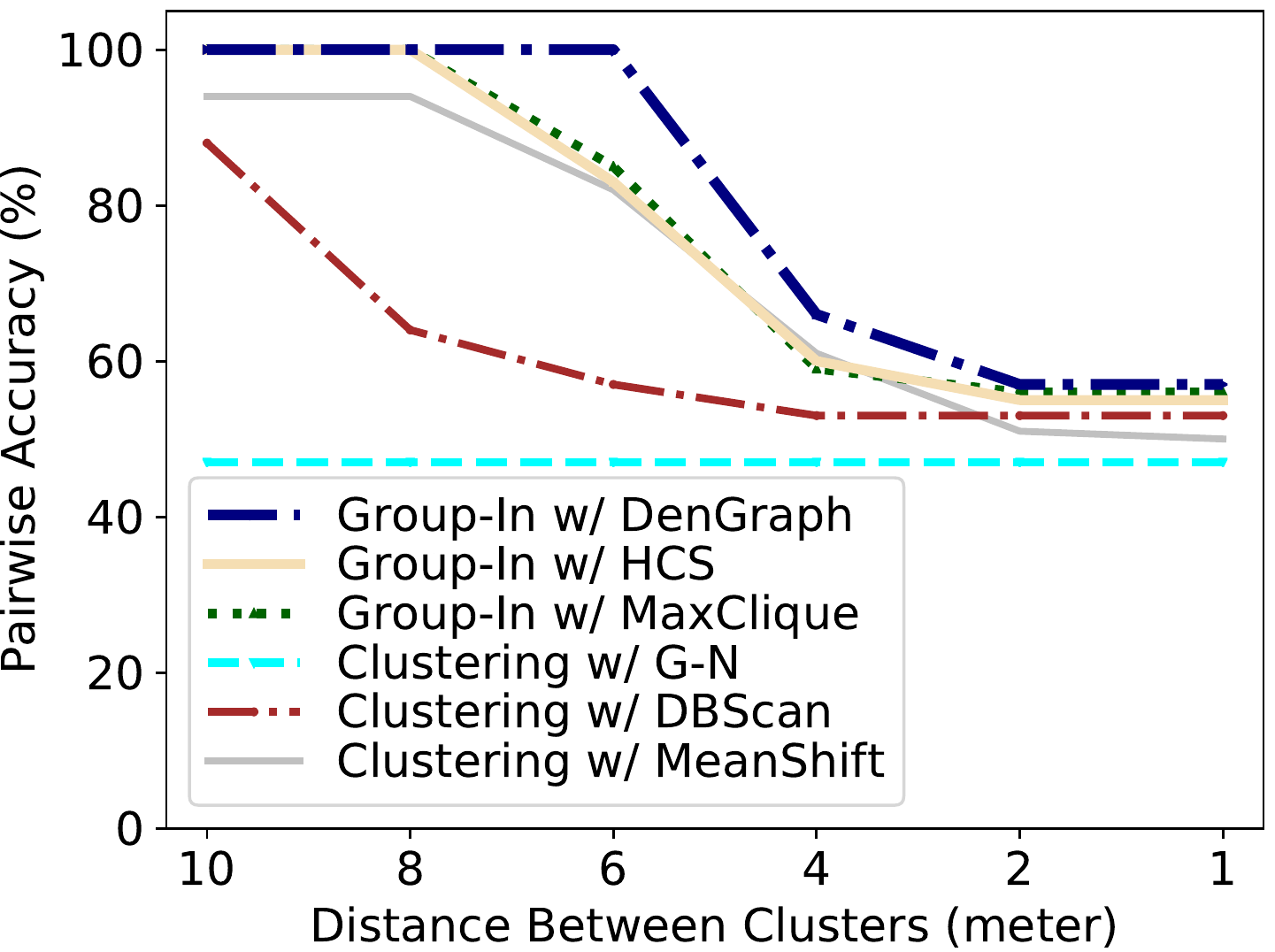}

    \end{subfigure}
    
    \begin{subfigure}[b]{0.22\textwidth}
        \includegraphics[width=1.01\textwidth]{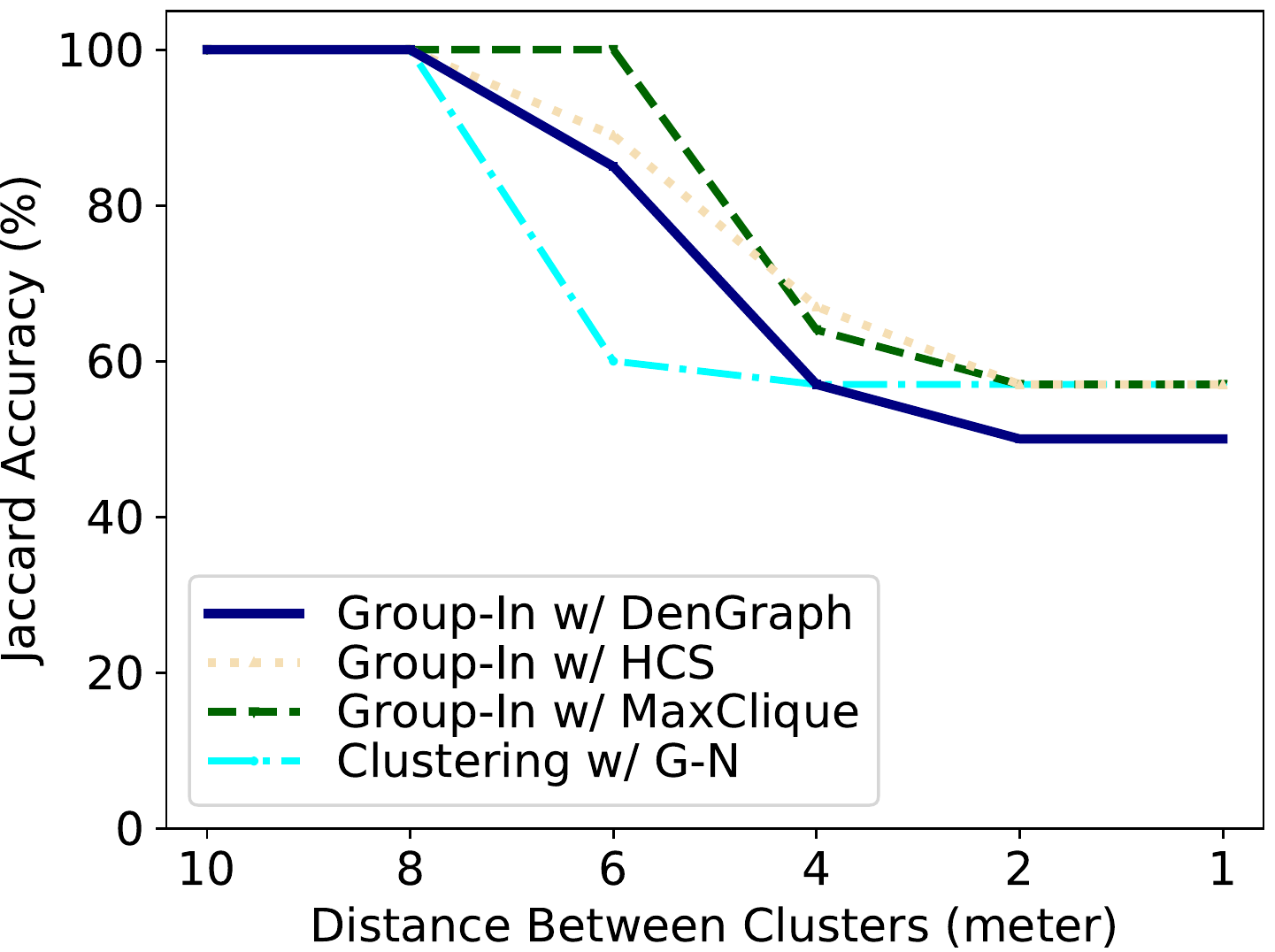}

    \end{subfigure}
    ~
        \begin{subfigure}[b]{0.22\textwidth}
        \includegraphics[width=1.01\textwidth]{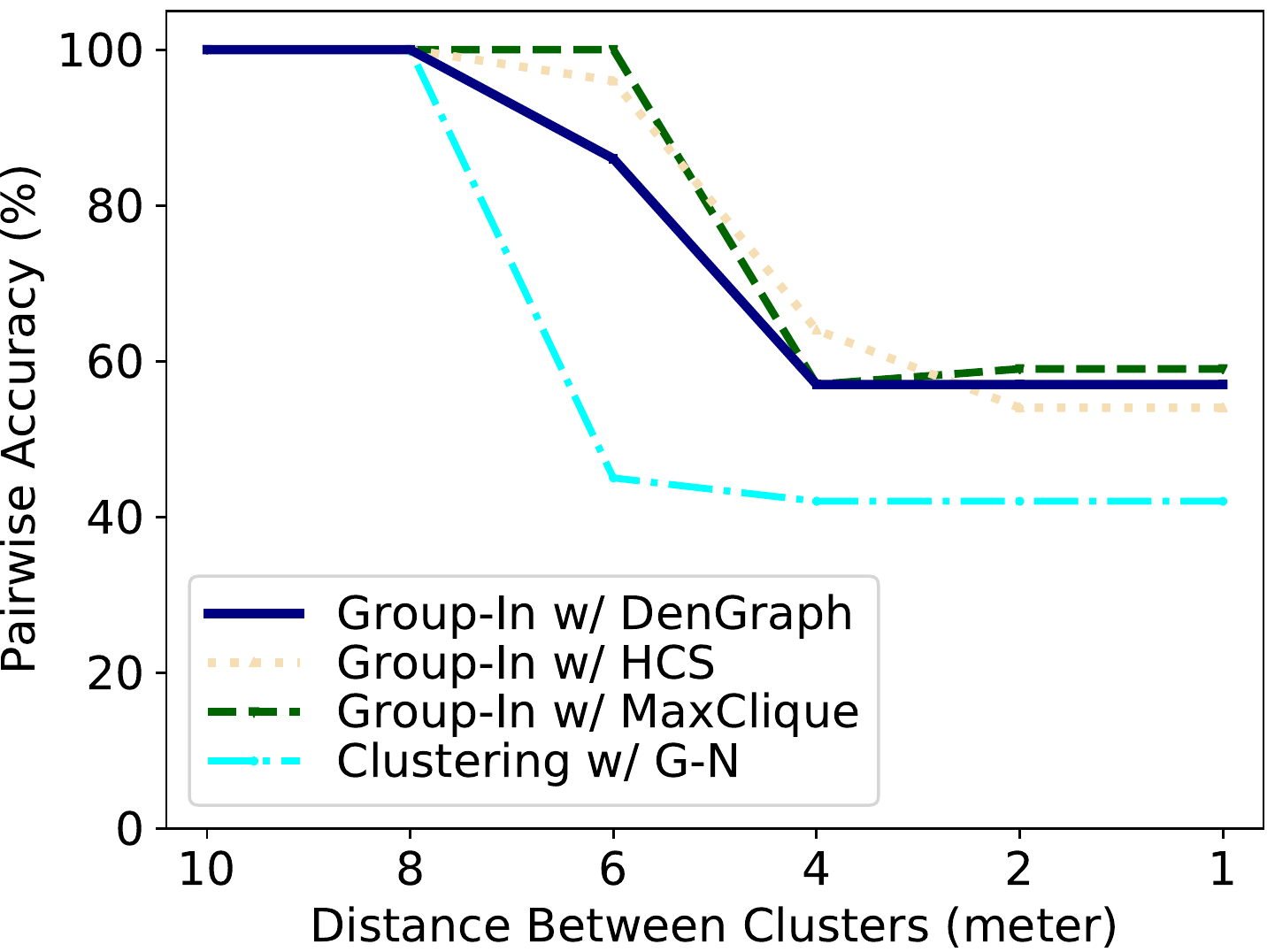}
    \end{subfigure}
  
    \caption{The precision results. Top: centralized, bottom: decentralized.}\label{DistanceExp}
\end{figure}

\noindent
\textbf{Results from the mobile scenarios:} Our second set of results target the evaluation of Group-In for mobile scenarios where the groups move frequently. In the first mobile scenario, we divide the beacons into two groups, and each group of beacons is carried by a person who walks straight (back and forth movements) between two corners of a square-shaped 100~sqm room. Each person starts from the opposite corner and walks with a similar walking pace (about 2~m/sec). The results in Fig.~\ref{fig:MobileExpStraight} show that centralized computing with the graph clustering algorithms achieves up to 98\% accuracy with 5~sec optimal sampling time. MaxClique achieves the highest Jaccard accuracy. On the other hand, direct clustering (DBScan and MeanShift) fails to detect mobile groups. Besides, decentralized computing is not able to capture the movement of groups, as it can only perform up to 70\% pairwise accuracy for 1~sec and 5~sec sampling times. The low accuracy is a result of single scanners not being able to differentiate two people walking back and forth in the same room on the same straight line (e.g., 10m) in short durations. In more extended and more realistic scenarios, decentralized computing may result in better accuracy. 

In the second mobile scenario, we two groups of people walk randomly in the room (based on Random Waypoint Model~\cite{bettstetter2004stochastic}). As shown in Fig.~\ref{fig:MobileExpRandom}, the accuracy of the second scenario is slightly lower as this mobile scenario is more dynamic than the previous (straight-walking) scenario. However, the results are still consistent with the previous scenario. For centralized computing, the graph clustering algorithms have higher accuracy than the direct clustering approaches and Girvan-Newman. Moreover, 5~sec sampling time is optimal, and it results in more than 80\% accuracy for both centralized and decentralized computing.

\noindent
\textbf{Precision results:} Our third set of results evaluate the precision of the proposed approach. We divide beacons into two groups and observe the effects of distance between the groups by gradually decreasing the distance from 10~m to 1~m. Fig.~\ref{DistanceExp} shows the accuracy w.r.t. the distances. Most of the clustering algorithms successfully detect groups in most cases if the distance is more than or equal to 4~m. In the case of very dense crowds, this may lead to a limitation if two groups always stay close to each other during the time interval.

\begin{figure}
\centering
    \begin{subfigure}[b]{0.47\textwidth}
      \includegraphics[width=1\textwidth,cfbox=black 0.5pt 0.5pt]{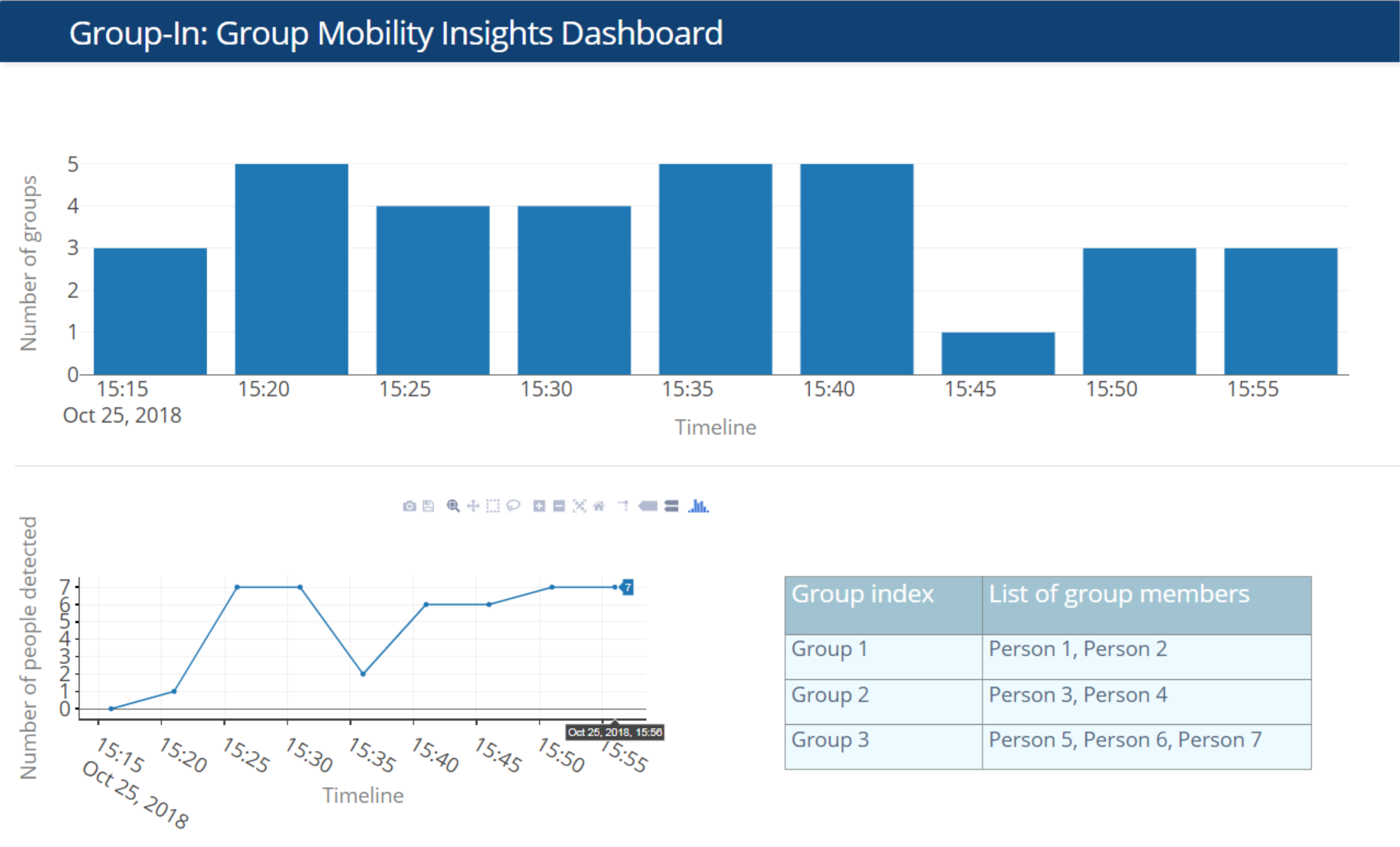}
\caption{A view from Group-In's live web interface for group inference.}
\label{Fig:WebInterface}

    \end{subfigure}\\
    \begin{subfigure}[b]{0.5\textwidth}
        
\includegraphics[width =0.9\textwidth]{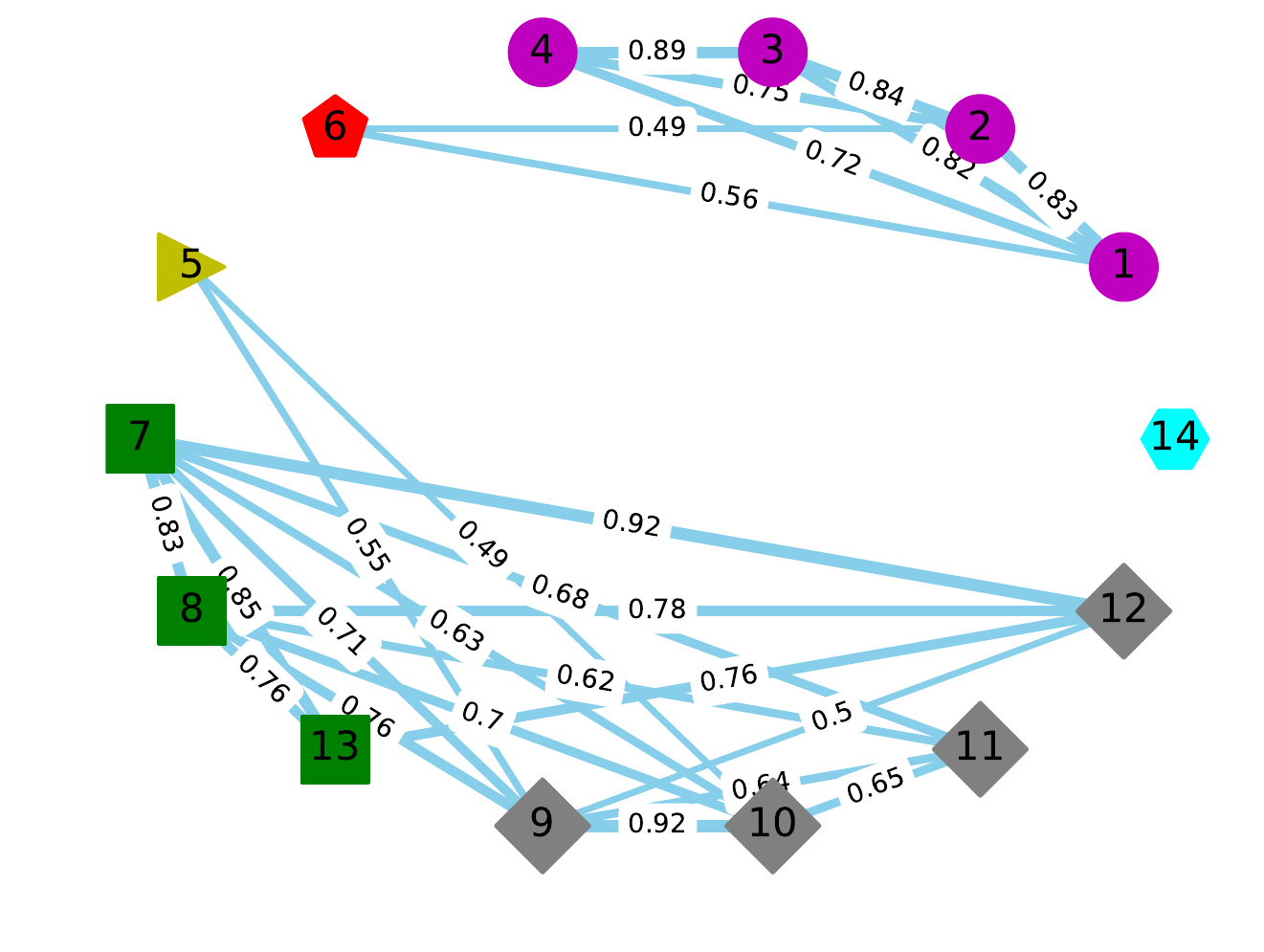}
\caption{Long-term linkage of 14 people from 1 month-long analysis.}
\label{fig:SocialNetwork}
\end{subfigure}
\caption{Group-In live dashboard and long-term linkage analysis results.}
\end{figure}

\noindent
\textbf{Visualizing group detection and long-term linkage results:} The results of the group detection are visualized in the Group-In's live web interface, as shown in Fig.~\ref{Fig:WebInterface}. Through this interface, group detection results such as the number of people, number of groups, and the sizes of each group can be monitored. The live interface provides real-time and offline visualization. 

The last set of results includes the long-term linkage evaluation in the real-world setup with the information of people's rooms, working groups, and their project groups during the experiment duration. In Fig.~\ref{fig:SocialNetwork}, the nodes representing people who share the same room have the same color and shape. The edge weights denote the linkage (see Section~\ref{Long-Term}). We observe that the linkage values of people who share the same rooms are mostly higher. In addition, Group-In is able to capture the relation of $P_6$ with $P_1$ and $P_2$. Although $P_6$ is a visitor member located outside of these four rooms, These three people belong to the same working group and spend time together. The other external member, $P_{14}$, is observed separately from this group. $P_{14}$ does not interact with other group members due to working in separate projects. For a better understanding of the physical proximity-based interactions, Fig.~\ref{fig:SocialNetwork} shows the edge weights of a person with others in different rooms,  whereas the people in the same rooms already have high linkage values due to being close to each other in the working place.

\noindent
\textbf{Remarks on the experiments:} Overall, Group-In can produce highly accurate group detection results in the short time intervals. The proposed approach successfully works with the sparse and noisy wireless data and changing number of dimensions. Moreover, although the experimented scenarios are very different from each other, we can apply the same set of parameters (without any training) and achieve high accuracy results in almost all scenarios. The only exception we observe is when the distance between clusters is persistently short (1 or 2~m). Lastly, the long-term linkage graph generated by Group-In can reflect the real conditions of the working environment considering room setups and project groups.

\noindent
\textbf{Limitations:} The limitations are observed mainly for decentralized computing. In particular, when the relatively stable mobility behaviors exist in a real-world office environment or the straight-walk scenario, the performance of the decentralized is significantly lower than the centralized computing. Furthermore, when two groups are too close to each other (e.g., 2~m apart) and not moving, the system may merge the two groups and regard them as only one group. 

We conducted the experiments using homogeneous BLE beacon devices. In the future applications, a possible limitation is the device heterogeneity~\cite{furst2018evaluating}. One can learn the characteristics of various devices and incorporate for improved accuracy in real-world setups where people use different wearables or smartphones. Moreover, multiple hardware from the same person (e.g., smartwatch and smartphone) can be considered in the future.

Lastly, the real-world experiment does not have very accurate ground-truth data. However, it has rather partial ground-truth data where the employees' working places and their expected movements are known. The long-term tracking of the people's movement for using cameras (e.g, body-worn cameras) can improve the accuracy results. At the same time, it may cause an invasion of privacy in an office environment. Therefore, we considered many controlled scenarios with different setups and group mobility behaviors. In one of the controlled scenarios, we conducted tests in the same environment by placing the devices in their expected rooms and observed that when employees are in their rooms, the system achieves close to 100\% accuracy. Although this alone does not prove high accuracy when employees do spontaneous daily movements, it indicates that in the case of ground-truth, the accuracy can be even higher.

%% file: Conclusion.tex
\section{Conclusion}

This paper proposes the Group-In system for group detection from wireless traces. Different from most indoor/outdoor localization approaches, which require extensive calibration efforts, Group-In does not aim high-accuracy localization and tracking of people's exact positions. On the other hand, it provides fast and accurate group detection results in real-time and offline. Moreover, the granularity of Group-In is better than existing group detection approaches, which assume that people can be in a group if the same scanner observes them. Our experiments in the lab scenarios and the real-world office environment provide confidence for Group-In's future usage in various urban environments such as campus environments, offices, museums, theme parks, and festivals.

\section{Acknowledgment}

This work has been funded by the EU Horizon 2020 Programme under Grant Agreements No. 731993 AUTOPILOT (Automated Driving Progressed by Internet Of Things) and No.871249 LOCUS (LOCalization and analytics on-demand embedded in the 5G ecosystem, for Ubiquitous vertical applicationS) projects. The content of this paper does not reflect the official opinion of the EU. Responsibility for the information and views expressed therein lies entirely with the authors.